%% file: a.tex
\documentclass[11pt]{article}

\usepackage{indentfirst}
\usepackage{outlines}
\usepackage{fixltx2e}
\usepackage{textcomp}
\usepackage{amsfonts}
\usepackage{verbatim}
\usepackage[english]{babel}
\usepackage{pifont}
\usepackage{color}
\usepackage{setspace}
\usepackage{lscape}
\usepackage{indentfirst}
\usepackage[normalem]{ulem}
\usepackage{booktabs}
\usepackage{natbib}
\usepackage{float}
\usepackage{latexsym}
\usepackage{url}

\usepackage{graphicx}
\usepackage{amssymb}
\usepackage{amsmath}
\usepackage{bm}
\usepackage{array}
\usepackage{mhchem}
\usepackage{ifthen}
\usepackage{hyperref}

\usepackage{caption}
\usepackage{xcolor}
\usepackage{amsthm}
\usepackage{amstext}

\input{macros-topic.tex}

\usepackage{algorithm, float}
\usepackage[noend]{algpseudocode}
\usepackage{setspace}

\newcommand{\blind}{0}

\newcommand{\aus}{s} 
\newcommand{\Aus}{\poset}

\newcommand{\bs}{{\color{black}\omega}} 
\newcommand{\Bs}{{\color{black}\Omega}}

\usepackage{soul}

\newcommand{\sls}{{\color{black}\zeta}}  

\begin{document}


\if0\blind
{
  \title{\bf Particle Gibbs Sampling for Bayesian Phylogenetic inference}
\author{Shijia Wang
   \hspace{.2cm}\\
   School of Statistics and Data Science, LPMC \& KLMDASR, Nankai University, China\\
    Liangliang Wang \\
   Department of Statistics and Actuarial Science, Simon Fraser University, BC, Canada\\
   \textbf{Contact:} Liangliang Wang, \href{lwa68@sfu.ca}{lwa68@sfu.ca}\\}  \maketitle
} \fi

\if1\blind
{
  \bigskip
  \bigskip
  \bigskip
  \begin{center}
    {\LARGE\bf Title}
\end{center}
  \medskip
} \fi

\abstract{The combinatorial  sequential Monte Carlo (CSMC) has been demonstrated to be an efficient complementary method to the standard Markov chain Monte Carlo (MCMC) for Bayesian phylogenetic tree inference using biological sequences.  It is appealing to combine the CSMC and MCMC in the framework of the particle Gibbs (PG) sampler to jointly estimate the phylogenetic trees and evolutionary parameters. However, the Markov chain of the particle Gibbs may mix poorly {\color{black} for high dimensional problems (e.g. phylogenetic trees). } 
Some remedies, including the particle Gibbs with ancestor sampling and the interacting particle MCMC,  have been proposed to improve the PG. But they either cannot be applied to or remain inefficient for the combinatorial tree space. 
 We introduce a novel  CSMC method by  proposing a more efficient proposal  distribution.  It also can be combined into the particle Gibbs sampler framework to infer parameters in the evolutionary model.  The new algorithm can be easily parallelized by allocating samples over different computing cores.
We validate that the developed CSMC can sample trees more efficiently in various particle Gibbs samplers  via numerical experiments. Our implementation is available at \href{https://github.com/liangliangwangsfu/phyloPMCMC}{https://github.com/liangliangwangsfu/phyloPMCMC} 
}

\maketitle

\section{Introduction}
The objective of phylogeny reconstruction methods is to recover the evolutionary history of biological species or other entries. A phylogenetic tree is latent, and typically estimated by using the biological sequences (e.g. DNA sequences) observed at tips of the tree. There is a rich literature on phylogenetic tree reconstruction.   
Bayesian approaches are extremely popular for phylogenetic inference. 
In Bayesian phylogenetics \citep{Lemey2009,drummond2010bayesian,huelsenbeck_mrbayes:_2001,Ronquist2003,Ronquist2012,Suchard15082006}, the goal of these methods is to compute a posterior over a  phylogenetic tree space. It is generally impossible to obtain an explicit expression for this posterior as the exact calculation involves integrating over all possible trees. The standard inference algorithm for Bayesian phylogenetics is Markov chain Monte Carlo (MCMC).  Many user-friendly software packages have been developed for implementing MCMC for
phylogenetic inference, such as MrBayes \citep{Ronquist2012}, BEAST \citep{suchard2018bayesian, bouckaert2019beast},
and BAli-Phy \citep{Suchard15082006}.

{\color{black} 
Sequential Monte Carlo (SMC) algorithms are popular for inference in state-space models \citep{Doucet01,liu01}, and can be applied to more general settings  \citep{del2006sequential}. 
There is a growing body of literature on phylogenetic tree reconstruction based on SMC methods.
Several SMC approaches \citep{teh08a,gorur09,Bouchard2012Phylogenetic,Gorur2012ScalableSMC} have been proposed 
to estimate clock trees and have been demonstrated to be good alternatives to MCMC methods.
These SMC approaches define the intermediate target distributions over forests over the observed taxa, and allow more efficient reuse of intermediate stages of the Felsenstein pruning recursions. 
A combinatorial sequential Monte Carlo proposed in \cite{LiangliangWang2015}  extends the previous work to construct both the clock and non-clock trees by correcting the bias in  the particle weight update in non-clock tree inference. \cite{LiangliangWang2015} also explore jointly estimating phylogenetic trees and parameters in evolutionary model in particle Metropolis Hastings framework. SMC algorithms have also been applied to online phylogenetic inference scenarios, in which the taxonomic data arrive sequentially in an online pattern \citep{dinh2017online, fourment2017effective}. \cite{dinh2017online} explore the theoretical property of their online SMC for phylogenetic inference. \cite{fourment2017effective} investigate the importance of proposal distributions for SMC in online scenarios. 
In addition, \cite{everitt2016sequential} has explored a combination of reversible jump methods with phylogenetic trees targeting the spaces of
varying dimensions. Several SMC algorithms for inference in intractable evolutionary models have been proposed \citep{Hajiaghayi2014Efficient, smith2017infectious}.  
\cite{Hajiaghayi2014Efficient} developed 
SMC methods for Bayesian phylogenetic analysis based on infinite state-space evolutionary models. \cite{smith2017infectious} develop SMC algorithm to jointly estimate phylogenetic tree and transmission network. 
An annealed sequential Monte Carlo  proposed in \cite{wang2018annealed}  can adaptively determine the sequence of intermediate target distributions in the general SMC framework \citep{del2006sequential}. 
}

\begin{figure}[htbp]
\centering
\includegraphics[width=0.7\textwidth]{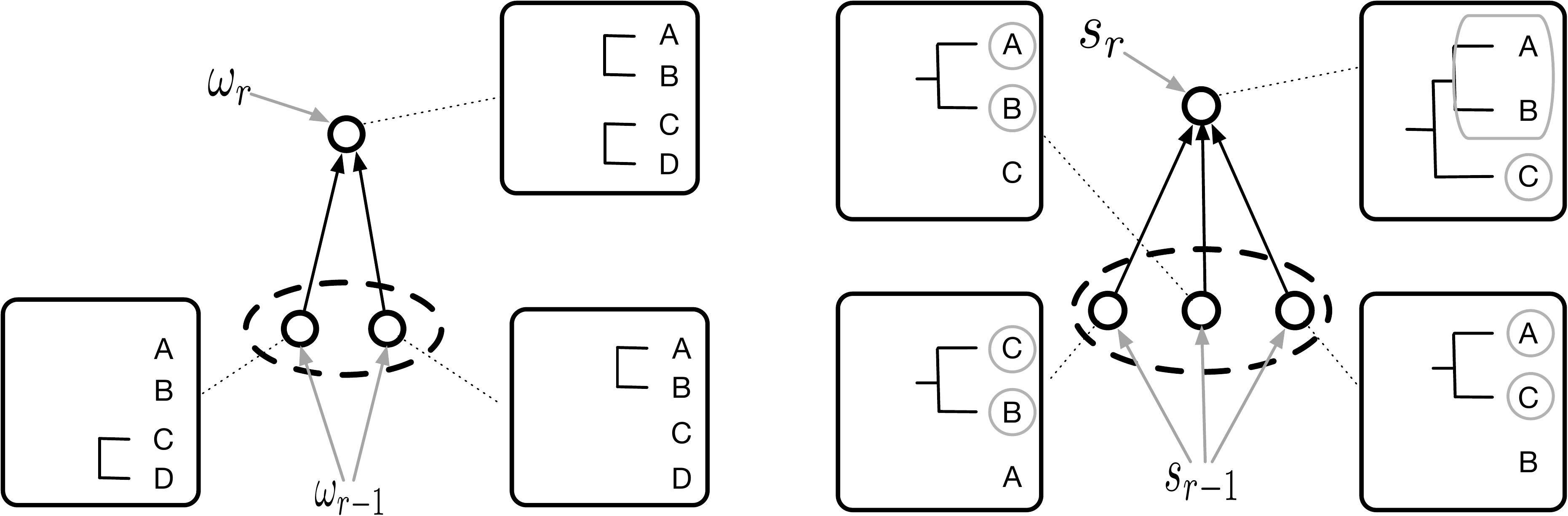}
\caption{Illustration of the overcounting issues in {\color{black} (a) original  CSMC.  The partial state $(A, B), (C, D)$ can be constructed in two different ways: by first merging $A$ and $B$, then merging $C$ and $D$, or by first merging $C$ and $D$, then merging $A$ and $B$; (b)  proposed CSMC. The augmented partial state $s_r$ can be constructed in three different ways. The two grey circles of each augmented partial state indicate the two subtrees that are just merged.}  } 
\label{fig:overcountingCSMC}
\end{figure}

In the combinatorial sequential Monte Carlo (CSMC)  \citep{LiangliangWang2015}, the proposal distribution can be more flexible than the one in \cite{Bouchard2012Phylogenetic} to propose non-clock trees. However, the standard  particle weight update cannot be applied to non-clock tree reconstruction because it will favour trees that can be constructed in multiple ways, which is called an overcounting issue, and therefore lead to biased estimates.  We provide an example to illustrate the overcounting issue in Figure \ref{fig:overcountingCSMC} panel (a).  This overcounting issue in non-clock tree inference  is corrected by introducing a backward kernel in CSMC.  This CSMC has been shown to be a good alternative  or a complementary method to MCMC for general Bayesian phylogenetics.

{\color{black} 

Particle Markov chain Monte Carlo (PMCMC)  \citep{Andrieu2010} is a general inference framework that combines SMC and MCMC, which are two standard tools for Monte Carlo statistical inference.  
However, path degeneracy limits the usage of SMC in approximating high dimensional targets. This issue arises due to the fact that the resampling step of SMC reduces the number of unique particles, as particles with large weights will be duplicated and particles with small weights will be pruned. 
This will lead to the results that SMC may use one or very few number of unique particle trajectories to approximate the target distribution. The path degeneracy issue can be mitigated by using a large number of particles, but this may induce a high computational cost. 
One type of PMCMC is the particle Gibbs (PG) sampler. PG iterates between sampling the static parameters and high dimensional latent variables (e.g. phylogenetic trees). 
  PG often suffers from a serious drawback that the mixing of the Markov chain can be poor when  the path degeneracy exists  in the underlying SMC. The underlying SMC may degenerate to a pre-specified reference trajectory, such that the latent variables may not be updated  through Gibbs iterations. Particle Gibbs with ancestor sampling (PGAs) was proposed in \cite{lindsten2014particle}  to enable fast mixing of the PG kernel even with a small number of particles in the underlying SMC  to reduce the computational burden. PGAs uses a so-called ancestor sampling (AS) step to update the reference trajectory. 
  Unfortunately, their proposed PGAs cannot be applied to the discrete tree space, which will be explained  in Figure 1 of \emph{Supplementary Material Section 5.1}.}

Our work is motivated by the need for an efficient  CSMC that is more robust to path degeneracy and can be utilized in the PGAs.  In Bayesian phylogenetics, the proposal distribution is important for exploring the complex tree posterior distribution. For instance, \cite{fourment2017effective} has investigated  different tree proposals in SMC and found that a good proposal is essential to exploring the posterior of trees. In this work, we focus on developing more efficient proposal distributions in SMC for the combinatorial space based on the CSMC in  \cite{LiangliangWang2015}.

We propose a novel CSMC algorithm with a novel proposal called CSMC-RDouP, which will be explained in Section \ref{sec:csmc-alg}.  The proposed method provides an easy framework of constructing a more flexible and efficient proposal based on a base proposal.  A backward kernel is proposed to correct the overcounting issue in CSMC-RDouP. The consistency properties of the estimators are guaranteed under weak conditions.  The CSMC-RDouP  is easy to parallelize by allocating samples into different  computing cores.  Further,  this new CSMC  can be combined with MCMC using various particle Gibbs samplers to  jointly estimate the phylogenetic tree and the associated evolutionary parameters. 
Our proposed method allows us to conduct ancestor sampling, which will be discussed later in the manuscript, to improve the mixing of PGs. 
We conduct a series of simulation studies to evaluate the quality of tree reconstruction using a variety of CSMCs and PGs.  Particle Gibbs with CSMC-RDouP  can estimate  trees more accurately than the one with a CSMC based on a base proposal. We also find that interacting particle Markov chain Monte Carlo \citep{rainforth2016interacting} is more efficient than particle Gibbs sampler with a fixed computational budget. 

\section{Background and notation}
\label{sec:bayesian-phylogenetic-inference}

We denote our observed biological sequence data by $y$. Let $X$ be a set of observed taxa. A phylogenetic X-tree $t$ represents 
the relationship among observed taxa via a tree topology and a set of branch lengths.   We only focus on the binary tree reconstruction. Let $\theta$ denote the parameter in a  nucleotide substitution model.  
The prior distribution of $\theta$ and $t$ are denoted by $p(\theta)$ and $p(t|\theta)$ respectively. The likelihood of data $y$ given $t$ and $\theta$ is denoted by $p(y|t, \theta)$. The joint posterior of $t$ and $\theta$ is denoted by $\pi(t, \theta)$. We introduce the notation $\gamma(\cdot)$ to denote the unnormalized posterior density.  

In Bayesian phylogenetics, our objective is to estimate the posterior distribution of $t$ and $\theta$, 
\begin{align}\label{eq:postTree}
\pi(\theta, t) = \frac{p(y|\theta, t)p(t|\theta)p(\theta)}{
p(y)}.
\end{align} 
Here $p(y)=\int \int p(y|\theta, t)p(t|\theta)p(\theta) \ud \theta \ud t $ is the marginal likelihood of biological sequence data.

With a site independence assumption, the likelihood function $p(y|\theta, t)$ can be evaluated by 
Felsenstein pruning \citep{felsenstein1973maximum, felsenstein1981evolutionary},   which involves the calculation of the probability of nucleotide mutation given a fixed amount of evolution (i.e. the branch length).  We use a continuous-time Markov chain (CTMC) to model the evolution of each site along each branch of $t$.
There is rich literature about phylogenetic nucleotide substitution models, such as  the Jukes-Cantor (JC) model \citep{jukes1969evolution}, the Kimura 2-parameter (K2P) model \citep{kimura1980simple}, and the general time reversible (GTR) model \citep{rodriguez1990general}.
The evolutionary model considered in this article is the K2P model. {\color{black} The rate matrix of the CTMC for K2P model only has one unknown parameter, $\kappa$, that represents the ratio of transition to transversion. In this case, the evolutionary model $\theta = \kappa$. See \emph{Supplementary Material Section 1} for details.}


A common  assumption in Bayesian phylogenetics is  that the priors for $\theta$ and $t$
are independent, i.e. $p(t|\theta) = p(t)$. A common prior over non-clock trees consists of a uniform distribution on topologies and a product of independent exponential distributions with rate $\lambda$ on branch lengths. A commonly used prior for  $\kappa$ in K2P model is an exponential distribution with rate $\mu_{0}$. We will use a coalescent tree prior for clock trees.

The exact evaluation of the normalized posterior $\pi(\theta, t)$ requires computing the marginal likelihood $p(y)$, which is generally intractable in phylogenetics.  We review classical MCMC methods for Bayesian phylogenetic inference in {\color{black} {\it Supplementary Material Section 2}} and list all notations in {\color{black} {\it Supplementary Material Table 2}}.

\section{Phylogenetic tree inference}
\label{sec:treeinference}
In this section, we assume that the parameter $\theta$ in the nucleotide substitution model is known. We are interested in the posterior inference over phylogenetic trees $\pi(t)$. 

\subsection{Combinatorial sequential Monte Carlo}
\label{sec:CSMC}
Combinatorial sequential Monte Carlo (CSMC) \citep{LiangliangWang2015} is an SMC algorithm for general tree inference based on a graded partially ordered set (poset) on an extended combinatorial space. 
The essential idea of the CSMC algorithm is to introduce a sequence of $R$ intermediate states to construct the target phylogenetic tree $t$ incrementally.  These $R$ intermediate states are typically graded from `simple' to `complex'. The CSMC algorithm sequentially approximates these intermediate distributions efficiently. The last intermediate distribution is the posterior of tree $\pi(t)$.

We use $\bs_1, \bs_2, \dots, \bs_R$ to denote the sequence of  intermediate states. We call state $\bs_r$ a partial state of rank $r$. For example, a partial state of rank $r$ can be a phylogenetic forest over the observed taxa, defined as a set of $R-r+1$ phylogenetic trees. We use the notation $\Bs_{r}$ to denote the set of partial states of rank $r$, and define $\Bs= \bigcup\Bs_{r}$. We use the notation $|\bs|$ to denote the number of trees in a forest $\bs$. 
Recall that our interest is in inferring $\pi(t)$ with $\gamma(t)$. Here $\gamma(t)$ denotes an unnormalized  $\pi(t)$. 
A natural extension from unnormalized posteriors on trees to the unnormalized posterior $\gamma$ on a forest is to take a product over the trees in the forest $\bs$ as follows: 
 \begin{align}\label{eq:natural-poset-extn}\gamma(\bs) = \prod_{(t_i, X_i)\in \bs} \gamma_{y(X_i)}(t_i), \end{align}
where $t_i$ denotes one tree $i$ in forest $\bs$, $X_i$ denotes the taxa of $t_i$ and $y(X_i)$ denotes the data associated with $X_i$.

CSMC algorithms iterate between resampling, propagation, and re-weighting to propose samples of rank $r$ from samples of rank $r-1$. We let $\bs_{r-1, k}$ denote the $k$-th particle of rank $r-1$. 
First, we resample $K$ times from the empirical posterior $ \pi_{r-1}(\bs)=\sum_{k = 1}^{K} W_{r-1, k}\delta_{\bs_{r-1,k}}(\bs)$, where $\delta$ is the delta function, and denote the particles after resampling by $\tilde{\bs}_{r-1, k}$, $k = 1, 2, \ldots, K$.
Second, we use a proposal distribution $\nu_{\tilde \bs_{r-1,k}}^+(\cdot)$ to propose samples $\bs_{r,k}$ from $\tilde \bs_{r-1,k}$.  
Finally, we compute a weight update for each particle. Figure \ref{fig:overcountingCSMC} (a) displays the overcounting issue for CSMC in non-clock tree inference. As shown in the figure, one intermediate state ($\bs_{r}$) may have multiple ancestors ($\bs_{r-1}$), which may lead to an inconsistent estimator if a standard SMC weight update is used. Instead, CSMC uses the following formula for the particle weight update: 
\begin{align}\label{eq:weight-update}
w_{r,k} &= \tilde w_{r-1,k}\cdot w(\tilde \bs_{r-1,k}, \bs_{r,k}), \\  \label{eq:weight-incremental-update}
 w(\tilde \bs_{r-1,k}, \bs_{r,k}) &= \frac{\gamma(\bs_{r,k})}{\gamma(\tilde \bs_{r-1,k})} \cdot  \frac{\nu_{\bs_{r,k} }^-( \tilde \bs_{r-1,k})}{\nu_{\tilde \bs_{r-1,k}}^+(\bs_{r,k})},
\end{align} 
where $\tilde w_{r-1,k}$ is the unnormalized particle weight from the previous iteration, $\nu^-$ is a backward kernel to correct an overcounting problem in non-clock tree inference and $\nu^+$ is the forward kernel that is the proposal  in the second step.  It is shown in  \cite{LiangliangWang2015} that a CSMC with the weight update in  (\ref{eq:weight-update}) can provide a consistent estimate of the posterior distribution.


\begin{figure}[htbp]
\centering
\includegraphics[width=0.5\textwidth]{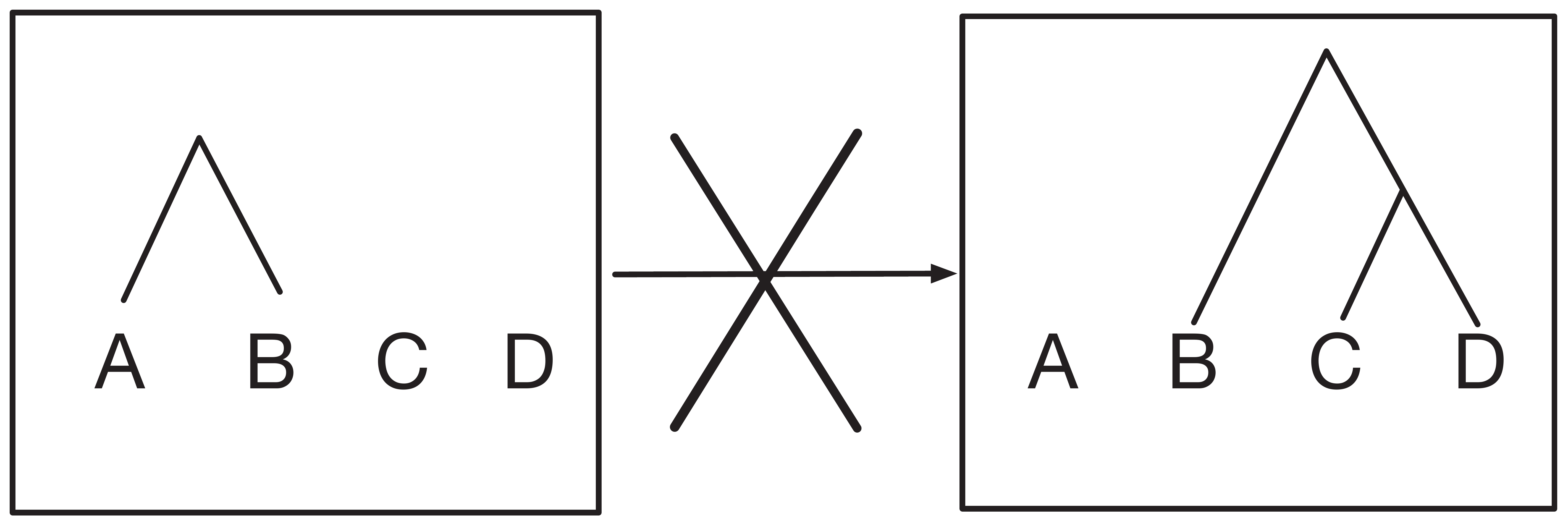}
\caption{An example to illustrate the limitation of the merge proposal. {\color{black} If a clade (A, B) exists in the partial state $\tilde{\bs}_{r-1}$, we cannot propagate clade (B, (C, D)) in the partial state $\bs_{r}$ using this merge proposal distribution. }}
\label{fig:CSMCweakness}
\end{figure}

The efficiency of the CSMC depends on several factors, including the choice of  proposal distributions and resampling schemes. This paper will only focus on the proposal distributions.  If the  proposal distribution is  inefficient, the path degeneracy issue of SMC can become  serious  when it is combined with MCMC in the framework of particle Gibbs. A  simple proposal is to propose new samples through randomly choosing a pair of subtrees to merge (i.e. picking a pair of trees in $\tilde{\bs}_{r-1}$ uniformly at random among the $\binom{|\tilde{\bs}_{r-1}|}{2}$ pairs) and sample new branch lengths. {\color{black} Note that these subtrees can also be singletons}. We will call this proposal \emph{the merge proposal} and denote $m^{+}_{\bs}(\bs')$ for the density of proposing $\bs'$ from $\bs$. This proposal is easy to implement but has some constraints. {\color{black} Figure \ref{fig:CSMCweakness} shows an example illustrating this constraint.  If a clade (A, B) exists in the partial state $\tilde{\bs}_{r-1}$, we cannot propagate samples of partial states $\bs_{r}$ without this clade. Consequently, we cannot propose $\{A, (B, (C, D))\}$ from $\{(A,B), C, D\}$.} This motivates us to develop a novel proposal distribution that can improve the performance of CSMC and can be used in particle Gibbs methods.

\subsection{CSMC with the RDouP proposal} 

\label{sec:csmc-alg}

We will improve the CSMC by constructing a novel and  sophisticated proposal based on its current proposal distribution.  To distinguish the two proposals and their corresponding CSMC algorithms, we will call the unimproved CSMC the vanilla CSMC and its proposal distribution a \emph{base propagation} or \emph{base proposal}.  Our new proposal distribution is based on the base proposal and will be called  the \emph{ RDouP proposal}, short for  Revert-Double-BasePropagation, for a reason which is explained later in this section.  Correspondingly, the CSMC with the RDouP proposal is named CSMC-RDouP.

{\color{black} The CSMC-RDouP algorithm will need a different sequence of intermediate states, denoted by $\aus_1, \aus_2, \ldots, \aus_R$. We call $\aus_r$ the $r$-th augmented (partial) state because it is based on the partial state in Section \ref{sec:CSMC}. Recall that a base partial state $\bs$ is a forest of trees with subtrees ${(t_i, X_i)\in \bs}, i=1,\ldots, |\bs|$. We let an augmented partial state $\aus$ be composed of a base partial state ${\color{black}\beta}(\aus)\in \Bs$ and some extra information related to the base proposal. With the merge base proposal, the extra information includes two trees that are children of one of the trees in ${\color{black}\beta}(\aus)$.

}

In CSMC-RDouP, we introduce $R$ intermediate target distributions {\color{black} on the $R$ augmented  states}. 
{\color{black} For the $r$-th augmented  state, $\aus_r$, we let $\gamma(\aus_r)=\gamma({\color{black}\beta}(\aus_r))$, which is the unnormalized posterior distribution for the forest ${\color{black}\beta}(\aus_r)$.  And $\aus_r$ is of rank $r$, the same rank as ${\color{black}\beta}(\aus_r)$. We use $\Aus_r$ to denote the set of augmented partial state of rank $r$, and define $\Aus=\bigcup\Aus_r$.}

\begin{figure}[htbp]
\centering
\includegraphics[width=0.7\textwidth]{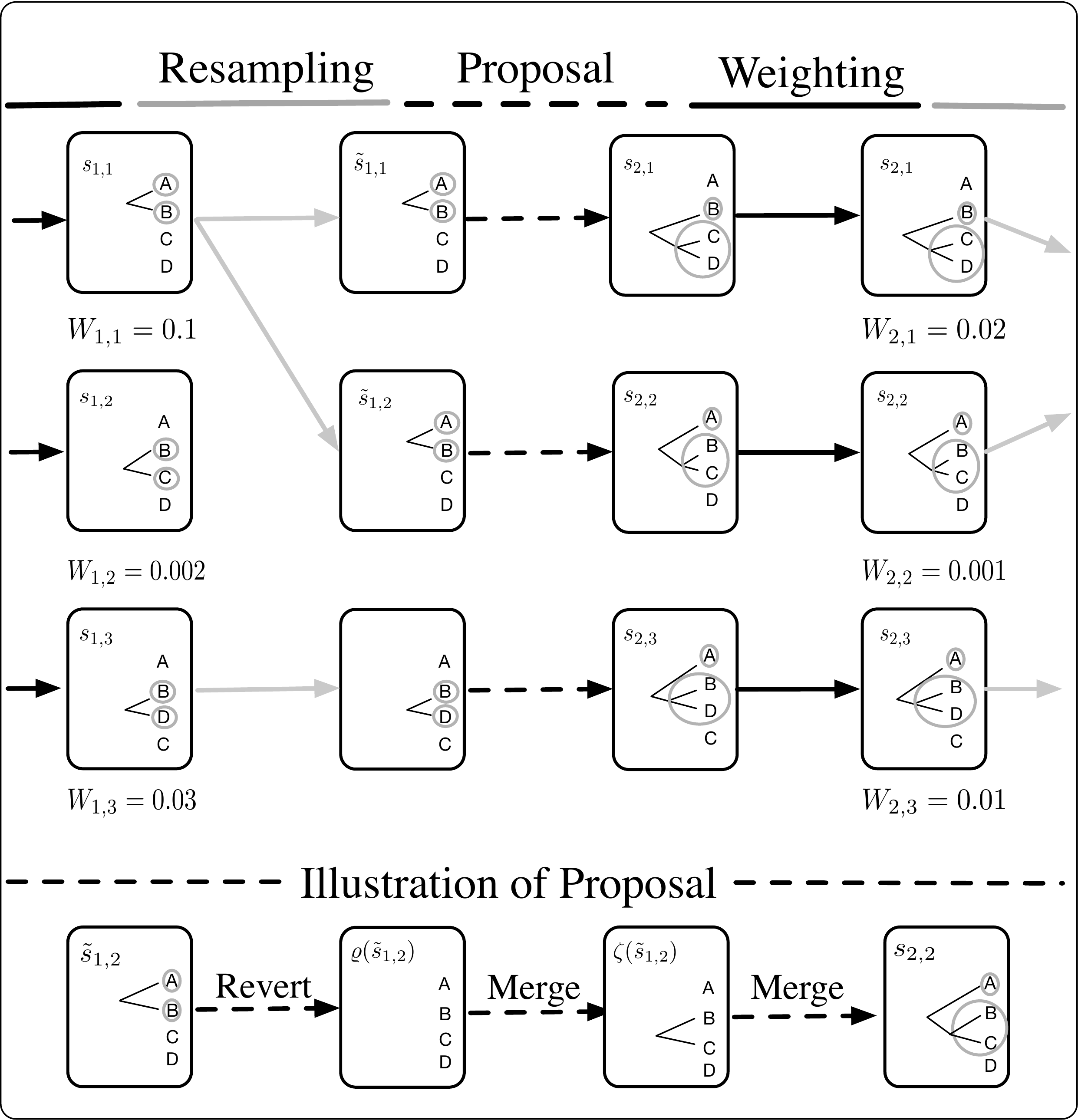}
\caption{ {\color{black} An overview of the CSMC-RDouP  framework. A set of partial states is kept at each SMC iteration. A positive-valued weight is associated with each partial state. Given a list of weighted particles of the partial state $\aus_{r-1}$, the CSMC-RDouP  algorithm performs the following three steps to approximate $\pi(\aus_{r})$: (i) resample to prune particles with small weights, (ii) propose a new partial state through the RDouP proposal, and
(iii) compute the weights for new particles. The lower panel of the figure provides an example of the proposal. The two grey circles of each particle indicate the two subtrees that are just merged. }}
\label{fig:SMCSMC}
\end{figure}

Figure \ref{fig:SMCSMC} presents an overview of the CSMC-RDouP  algorithmic framework. The CSMC-RDouP  algorithm sequentially approximates $\pi(\aus_{r})$ $(r = 2, 3, \ldots, R)$.  The algorithm is initialized at rank $r=1$ by initializing the list with $K$ copies of the least partial state $\aus_{1}$ (a list of taxa without any connections among them) {\color{black} and an empty set of trees that are most recently merged} with the same weight. Given a list of weighted particles of the partial state $\aus_{r-1}$, the CSMC-RDouP  algorithm performs the following three steps to approximate $\pi(\aus_{r})$:  resampling, propagation, and re-weighting.

{\bf Resampling:} First, we conduct a resampling step to resample $K$ particles from the empirical distribution $ \pi_{r-1}(\aus)=\sum_{k = 1}^{K} W_{r-1, k}\delta_{\aus_{r-1,k}}(\aus)$ and denote the resampled particles by $\tilde \aus_{r-1,1}$, $\tilde \aus_{r-1,2}$, $\ldots$, $\tilde \aus_{r-1,K}$. The resampling step prunes particles with low weights. A list of equally weighted samples is obtained after performing the resampling step.
{\color{black}
Instead of conducting resampling at every SMC iteration, we resample particles in an adaptive fashion \citep{Doucet11tutorial}. We compute a measure of
particle degeneracy at every iteration, and perform resampling only when the particle
degeneracy exceeds a pre-determined threshold. Effective sample size (ESS) is the standard criteria for measuring the particle degeneracy. 
 To make the algorithm more efficient, we only resample when the relative effective sample size (rESS) falls below a threshold. The relative effective sample size (rESS) is defined as s
$\text{rESS}(W_{\cdot}) = \bigg(K\sum_{k = 1}^{K}W_{k}^{2}\bigg)^{-1}$,
where $W_{\cdot}$ represents a vector of length $K$ for the normalized particle weights. }

{\bf Propagation:} Second, we propagate a new particle of rank $r$, denoted by $\aus_{r,k}$ from each of the resampled particle $\tilde \aus_{r-1,k}$ ($\tilde \aus_{r-1,k} = \aus_{r-1,k}$ if we do not conduct resampling at rank $r-1$),  using a proposal distribution $\nu^+_{\tilde \aus_{r-1,k}} : \Aus \to [0, 1]$. 

We construct a sophisticated proposal based on a base proposal using three steps. The first step is to undo the last base proposal; the second step is to  conduct one base proposal; and the third step is to do another base proposal {\color{black} and keep relevant information for undoing this move later}. In other words, the new proposal is based on reverting the {\color{black} augmented}  partial state and then double implementing the base propagation, called  Revert-Double-BasePropagation (RDouP) proposal.

To be more concrete, we will introduce some notation and use the merge proposal as an example of the base proposal. {\color{black} The reverted state of an augmented partial state, $\aus =\{\bs, (t_i, X_i), (t_j, X_j)\}$, is obtained by removing the tree with two children $(t_i, X_i)$ and  $(t_j, X_j)$ from  $\bs$, but keeping $(t_i, X_i)$ and  $(t_j, X_j)$. That is, 
 \begin{align}\label{eqn: reverted state}
  \varrho(\aus) = \{\bs\backslash ((t_i, X_i), (t_j, X_j))\} \cup \{(t_i, X_i), (t_j, X_j)\},   
 \end{align}
 where $((t_i, X_i), (t_j, X_j))$ is the clade with children $(t_i, X_i)$ and $(t_j, X_j)$. In contrast, we can reduce an augmented partial state to a base partial state
 \begin{align}\label{eqn: beta of aus}
 {\color{black}\beta}(\aus) = \bs = \aus\backslash  \{(t_i, X_i), (t_j, X_j)\}.   
 \end{align}
}
 The probability density of proposing a new state $\aus_{r,k}$ from the current state $\tilde \aus_{r-1,k}$ is denoted by $\nu^+_{\tilde \aus_{r-1,k}}(\aus_{r,k})$.
There are 3 steps in propagating $\aus_{r,k}$:
\begin{enumerate}
\item Find the reverted  {\color{black} base}   state of the current {\color{black} augmented} partial state $\tilde \aus_{r-1,k}$, denoted by $\varrho(\tilde \aus_{r-1,k})$. 
\item We first pick
a pair of trees in $\varrho(\tilde \aus_{r-1,k})$ uniformly at random among the $\binom{|\varrho(\tilde \aus_{r-1,k})|}{2}$ pairs, and then sample the length of the new branches. This state is denoted by $\sls(\tilde \aus_{r-1,k})$. 

\item We pick
a pair of trees in $\sls(\tilde \aus_{r-1,k})$ uniformly at random among the $\binom{|\sls(\tilde s_{r-1,k})|}{2}$ pairs, and sample the length of the new branches.  {\color{black} We also keep the two trees that are merged together with the progagated partial state.} This step finishes the propagation of $\aus_{r, k}$.
\end{enumerate}
To ease the description of the algorithm, we call $\tilde{\aus}_{r-1, k}$ the parent state of $\aus_{r, k}$, and call $\varrho(\tilde \aus_{r-1,k})$ the reverted state of $\tilde \aus_{r-1,k}$.  {\color{black} The lower panel of Figure \ref{fig:SMCSMC} displays an example of proposing $\aus_{r,k}$ from $\tilde \aus_{r-1,k}$ using the RDouP proposal based on the merge proposal. In this example, we first find the reverted state of $\{(A, B), C, D), A, B\}$, which is $\{A, B, C, D\}$. Then we merge $B$ and $C$,  and merge $(B, C)$ and $A$ to form the augmented partial state, $\{(A, (B, C)), D, A, (B, C)\}$. }
The parent state of $\aus_{r, k}$ is $\tilde{\aus}_{r-1, k}$, and its reverted state is $\sls(\tilde \aus_{r-1,k})$.  
s

The forward kernel in CSMC-RDouP  has the following form: 
\begin{align} 
\label{eqn:forward}
 \nu_{\tilde \aus_{r-1,k}}^+(\aus_{r,k})  =  r_{\tilde \aus_{r-1,k}}^{+}({\varrho}(\tilde \aus_{r-1,k})) \cdot  m_{{\varrho}(\tilde \aus_{r-1,k})}^{+}( \sls(\tilde \aus_{r-1,k})) \cdot  {\color{black} \check{m}}_{\sls(\tilde \aus_{r-1,k})}^{+}({\color{black} \aus_{r,k}}), 
\end{align}
where $r^{+}_{\aus}(\bs)$ is 1 if $\bs$ is the reverted state of $\aus$, otherwise 0;   $m^{+}_{\bs}(\bs')$ is the density of proposing {\color{black} a base state} $\bs'$ from  {\color{black} a base state} $\bs$;  {\color{black}  $\check{m}^{+}_{\bs}(\aus)$ is the density of proposing an augmented state $\aus$ from  a base state $\bs$. {\color{black} Note that in $\check{m}_{\bs}^+(\aus)$, we propose an augmented state $\aus$ from a base state $\bs$ by keeping the two trees that are most recently merged; hence, $\check{m}^{+}_{\bs}(\aus) = m_{\bs}^+{(\beta(\aus))}$,  where $\beta(\aus)$ is a base state reduced from the augmented state $\aus$.} }

{\bf Re-weighting:} Finally, we compute a weight for each of these new particles using the same weight update formula in Equation (\ref{eq:weight-update}) in the vanilla version of CSMC because the RDouP proposal also generates the  overcounting issue.  Figure \ref{fig:overcountingCSMC} (b) illustrates the overcounting issue in CSMC-RDouP, where there are multiple ways to propose the same {\color{black} augmented} partial state. 
Due to the overcounting issue,  a backward kernel is required in the weight update to obtain a consistent estimate for the posterior distribution.

We propose to use the backward kernel as follows: 
\begin{align} 
\label{eqn:backward}
& \nu_{\aus_{r,k}}^-(\tilde \aus_{r-1,k})  \\
= &  m_{{\color{black}\beta(\aus_{r,k})}}^{-}(\sls(\tilde \aus_{r-1,k}))  \cdot  m_{\sls(\tilde \aus_{r-1,k})}^{-}(\varrho(\tilde \aus_{r-1,k})) \cdot  r_{ \varrho(\tilde \aus_{r-1,k})}^{-}(\tilde \aus_{r-1,k}) \nonumber,
\end{align}
where 
$m^{-}_{\bs}(\bs')>0$ if there are multiple ways proposing  a base state $\bs'$ from a base state $\bs$ using $m^{+}$, and $r^{-}_{\bs}(\aus)>0$  if $\bs$ is the reverted state of {\color{black}the augmented state} $\aus$. We will show that this choice of backward kernel  can lead to asymptotically consistent estimates in the {\color{black} \emph{Supplementary Material Section 3}}.

By plugging  (\ref{eqn:forward}) and (\ref{eqn:backward}) into (\ref{eq:weight-incremental-update}), the incremental weight  function can be rewritten as 
\begin{align} 
&w(\tilde \aus_{r-1,k}, \aus_{r,k}) \label{eqn:RDouP-IncrementalWeight}\\ 
= & \frac{\gamma(\aus_{r,k})}{\gamma(\tilde \aus_{r-1,k})} \cdot  \frac{m_{{\color{black}\beta(\aus_{r,k})}}^{-}(\sls(\tilde \aus_{r-1,k}))  \cdot  m_{\sls(\tilde \aus_{r-1,k})}^{-}(\varrho(\tilde \aus_{r-1,k}))\cdot r_{ \varrho(\tilde \aus_{r-1,k})}^{-}(\tilde \aus_{r-1,k})}{ r_{\tilde \aus_{r-1,k}}^{+}({\varrho}(\tilde \aus_{r-1,k})) \cdot m_{{\varrho}(\tilde \aus_{r-1,k})}^{+}( \sls(\tilde \aus_{r-1,k})) \cdot  m_{\sls(\tilde \aus_{r-1,k})}^{+}({\color{black}\beta(\aus_{r,k})})} \nonumber.
\end{align}

We construct a discrete positive measure using a list of weighted particles at rank $r$, $\label{eq:measure-approximation}\pi_{r,K}(\aus) = \sum_{k=1}^K W_{r,k} \delta_{\aus_{r,k}}(\aus), \textrm{for all } \aus \in \Aus$. 
In the end, we obtain a Monte Carlo approximation $\pi_{R,K}$ of  $\pi(t)$.  
Algorithm~\ref{algo:csmc} summaries the CSMC-RDouP  algorithm.

\begin{algorithm}
\caption{: \textbf{RDouP combinatorial sequential Monte Carlo}} 
\label{algo:csmc}
  {\fontsize{11pt}{11pt}\selectfont
\begin{algorithmic}[1]
  \State {\bfseries Inputs:} (a) Prior over {\color{black} augmented} partial states $p(\aus)$;  
(b) Likelihood function $p(y | \aus, \theta)$; (c) Threshold of the rESS: $\epsilon$.

   \State {\bfseries Outputs:}  Approximation of the posterior distribution, $\sum_k W_{R,k} \delta_{ \aus_{R,k}}(\cdot) \approx \pi(\cdot)$.
	\State  Initialize SMC iteration index: $r \leftarrow 1$.  

 \For{$k\in\{1, \dots, K\}$}
  \State Initialize particles with the least partial state. 
  \State  Initialize  weights: $w_{1,k} \leftarrow 1$; $W_{1,k} \leftarrow 1/K$.
  \EndFor

\For{ rank $r \in\{2, \dots, R\}$}
\If{$\text{rESS}( W_{r-1,\cdot}) < \epsilon$} 
\State Resample the particles. 
   \For{$k \in \{1, \dots, K\}$}
    \State \label{step:adapt-reset-weights} Reset particle weights: $\tilde w_{r-1,k} = 1$; $\tilde W_{r-1,k} = 1/K$.
    \EndFor
    \Else
    \For{$k \in \{1, \dots, K\}$}
    \State $\tilde w_{r-1,k} = w_{r-1,k} ;\tilde \aus_{r-1,k} =  \aus_{r-1,k} $.     \EndFor
\EndIf
 \For{ all $k\in\{1, \dots, K\}$}

       \State Sample particles $ \aus_{r,k} \sim \nu^+_{ \tilde \aus_{r-1,k}}(\cdot)$, using one revert move to find $\varrho(\tilde \aus_{r-1,k})$, and two base proposals to propose $\sls(\tilde \aus_{r-1,k})$ and $\aus_{r,k}$. 
         \State Update weights according to Equation (\ref{eq:weight-update}). 
  \EndFor
  \EndFor
\end{algorithmic} 
}
 \end{algorithm}

\noindent   {\assumption
 For all $s$, $s'\in\poset$, $m^{+}_{s}(s') = 0$ implies $m^{-}_{s'}(s) = 0$.}
 
 \setcounter{theorem}{0}

\noindent  {\proposition  For all $\aus$, $\aus'\in\Aus$, if $m^{+}$ and $m^{-}$ satisfy Assumption 1,  $\nu^{+}_{\aus}(\aus') = 0$ implies $\nu^{-}_{\aus'}(\aus) = 0$. }

\noindent {\proof
We have $\nu_{\aus}^+(\aus') = r^{+}_\aus(\varrho(\aus)) \cdot m_{\varrho(\aus)}^{+}(\sls(\aus)) \cdot  m_{\sls(\aus)}^{+}({\color{black}\beta(\aus')})$.  Since $\varrho(\aus)$ is the reverted state of $\aus$,  $r^{+}_\aus(\varrho(\aus))=1$. 
Hence 
$\nu_{\aus}^+(\aus') =   0$  implies that either 
$m_{\varrho(\aus)}^{+}(\sls(\aus))=0$ or $m_{\sls(\aus)}^{+}({\color{black}\beta(\aus')})$ is 0.   

Based on the Assumption 1 on the base proposal, $m_{\varrho(\aus)}^{+}(\sls(\aus))$ implies $m_{\sls(\aus)}^{-}(\varrho(\aus)) =0$, and  $m_{\sls(\aus)}^{+}({\color{black}\beta}(\aus'))$ implies $m_{{\color{black}\beta}(\aus')}^{-}(\sls(\aus))$=0.  
Hence, either $m_{\sls(\aus)}^{-}(\varrho(\aus)) =0 $ or $m_{{\color{black}\beta}(\aus')}^{-}(\sls(\aus))=0$.  Since $\nu_{\aus'}^-(\aus) =  m_{{\color{black}\beta}(\aus')}^{-}(\sls(\aus))  \cdot  m_{\sls(\aus)}^{-}(\varrho(\aus)) \cdot r^{-}_{\varrho(\aus)}(\aus)$, we have $\nu_{\aus'}^-(\aus)=0$. }

\noindent { \proposition If for all $\aus$, $\aus'\in\poset$, $\nu^{+}_{\aus}(\aus') = 0$ implies $\nu^{-}_{\aus'}(\aus) = 0$, the CSMC-RDouP  provides asymptotically consistent estimates. We have
\[
\sum_{k = 1}^{K}W_{R, k}\phi(\aus_{R,k}) \to \int \pi_{R,k}(\aus)\phi(\aus)\text{d}\aus ~~\text{as}~~K\to \infty,
\]
where the convergence is in $L^{2}$ norm,  and $\phi$ is a target function under mild conditions. For example, $\phi$ is a bounded function.}

\noindent  {\color{black} By construction, the RDouP proposal will induce a ranked poset defined on $\Aus$ such that $\aus'$ covers $\aus$ if and only if $w(\aus, \aus')>0$. In this poset structure, an augmented partial state $\aus$ is deemed to precede another augmented partial state $\aus'$ if $\aus'$ can be reached by obtaining the reverted state of $\aus$ followed by conducting two times of the base proposal and keeping the two trees that are merged in the second merge proposal. Consequently, the proposed CSMC-RDouP is within the framework of CSMC in \cite{LiangliangWang2015} and the consistency of posterior estimates is  guaranteed by Proposition 2 in it. We refer readers to \cite{LiangliangWang2015} for the proof of consistency.}

{\color{black} In the \emph{Supplementary Material Section 3}, we explain the weight update for  clock trees and non-clock trees in detail.}

\section{Joint estimation of phylogenetic tree and evolutionary parameter}
\label{sec:PGS}
\subsection{Particle Gibbs Sampler}
In a more realistic scenario, $\theta$ is also an unknown parameter that requires us to estimate given data. 
We study the particle Gibbs sampler, a Gibbs-type algorithm that iterates between sampling $t$ and $\theta$. Given a tree $t$, we use one Metropolis-Hastings step to sample $\theta$ from $\pi(\theta|t)$. 
Given $\theta$, a conditional CSMC-RDouP algorithm (described in \emph{Supplementary Material Section 4})  is used to approximate $\pi(t|\theta)$.  The main difference between the CSMC-RDouP algorithm and the conditional version is that in the latter, one of the particle trajectories is pre-specified, which is called the reference trajectory. This reference trajectory cannot be pruned in the resampling step. The resulting Markov chain of PGs will leave the target distribution invariant for an arbitrary number of particles used in the conditional CSMC-RDouP. 
Without loss of generality, we assume the first particle trajectory $\aus_{1:R, 1}$ to be the reference trajectory, denoted by $\aus_{1:R}^{*}$. In PGs, we first use one Metropolis-Hastings step to update the parameter $\theta$, and then conditional on this $\theta$, we sample a particle trajectory from the approximated posterior of phylogenetic forests by running the conditional CSMC-RDouP. This sampled trajectory will be the reference trajectory of the conditional CSMC-RDouP in the next particle Gibbs iteration. We iterate these two steps until the convergence is achieved. We summarize the algorithm of PGs in the 
\emph{Supplementary Material Section 4}.

\subsection{Particle Gibbs Sampler with Ancestor Sampling}
\label{section:Particle Gibbs Sampler with Ancestor Sampling}

In particle Gibbs, the reference trajectory is kept intact throughout the CSMC-RDouP  algorithm. This may lead to slow mixing of the PGs algorithm when path degeneracy exists. We investigate particle Gibbs with ancestor sampling (PGAs) \citep{lindsten2014particle}  to improve the mixing of particle Gibbs samplers. The basic idea of PGAs is to include an ancestor sampling step in the conditional CSMC-RDouP  algorithm to update the reference trajectory. 
If the reference trajectory is updated through the ancestor sampling step, the particle system may degenerate to a new trajectory other than the reference trajectory. {\color{black} We illustrate the  implementation of the ancestor sampling step in the \emph{Supplementary Material Section 5.1}.}

\subsection{Interacting particle Markov chain Monte Carlo}
Another type of the particle Gibbs algorithm, interacting particle Markov chain Monte Carlo (IPMCMC) \citep{rainforth2016interacting}, is considered to improve the mixing of a particle Gibbs sampler. In IPMCMC, a pool of  standard and conditional  CSMC-RDouP  algorithms are interacted to design an efficient proposal for tree posterior. The interaction of conditional and standard CSMC-RDouP algorithms is achieved by communicating their marginalized likelihoods. The algorithmic description of  IPMCMC is displayed in the  {\color{black} \emph{Supplementary Material Section 5.2}}. 
The standard and conditional CSMC-RDouP  can be allocated into different computing cores to achieve parallelization.

\section{Simulation studies}
\label{sec:psimulation}

{\color{black} We assess the performance of the CSMC-RDouP method  with simulation studies and provide main findings in this section. Please refer to \emph{Supplementary Material Section 6} for details.}

We first emphasize a comparison of the vanilla CSMC and the proposed CSMC-RDouP in terms of computational speed.  We find that the relative runtime of CSMC-RDouP  compared with CSMC is about 1.4.  The computational speed of CSMC-RDouP  is lower than that of the vanilla  CSMC. This is expected as the proposal in the vanilla CSMC is simpler, while in our new proposal we have to use one move to find the reverted state and merge twice to propose the new partial state. The weight update function in CSMC-RDouP  algorithm is also more complicated.

\begin{figure}[htp]
\centering
\includegraphics[width=0.8\textwidth]{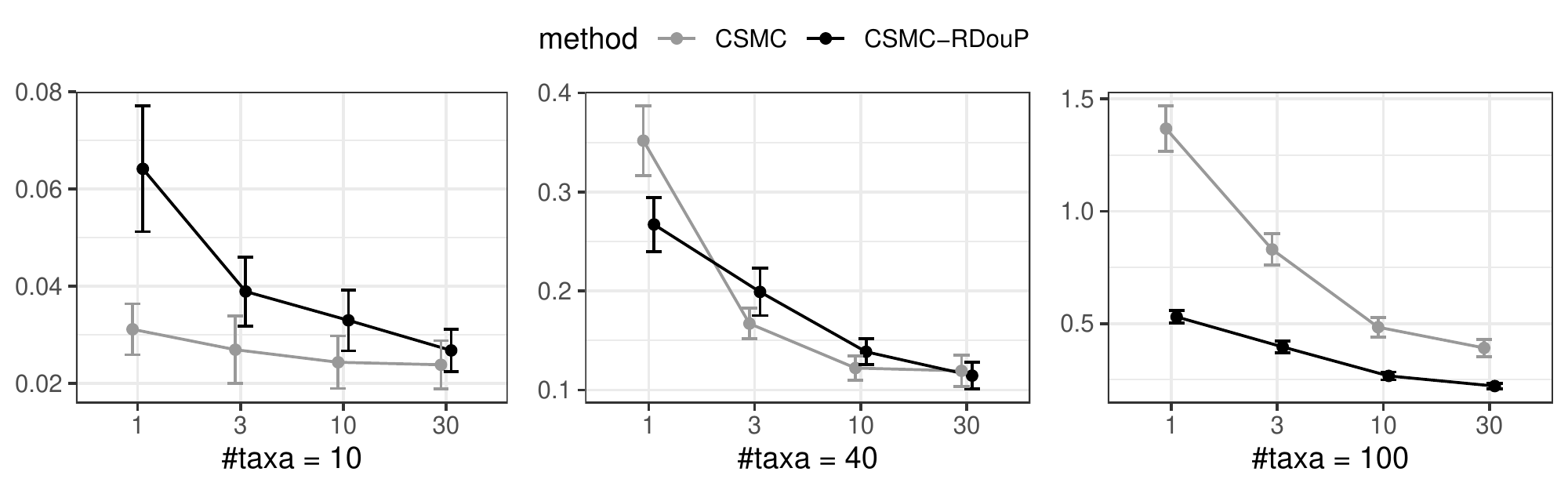}
\caption{{\color{black} Comparison of CSMC and CSMC-RDouP  as a function of number of (1000) particles in three scenarios: $10$ taxa (left), $40$ taxa (middle), $100$ taxa (right).  The x-axis represents the number of a thousand particles.  The four levels of $K$ (number of particles)  from left to right of each panel for  CSMC-RDouP are $1\cdot 10^{3}$, $3\cdot 10^{3}$, $1\cdot 10^{4}$ and $3\cdot 10^{4}$ respectively. The four levels of $K$ from left to right of each panel for vanilla version of  CSMC are $1.5\cdot 10^{3}$, $4.5\cdot 10^{3}$,  $1.5\cdot 10^{4}$, $4.5\cdot 10^{4}$ respectively. The y-axis represents the Robinson Foulds metric. }}
\label{fig:CSMC-RM}
\end{figure}

We also compare the vanilla CSMC and the proposed CSMC-RDouP in terms of the tree reconstruction quality.  The majority-rule  consensus tree is used to  summarize the weighted samples of phylogenetic trees \citep{felsenstein1981evolutionary}.  We  use the Robinson-Foulds (RF) metric based on sums of differences in branch lengths metric \citep{RobinsonFoulds1979} to measure the distance between the estimated trees and true trees. From Figure \ref{fig:CSMC-RM}, we conclude that  the tree reconstruction accuracy increases with incremental numbers of particles. For larger trees, the reconstruction quality provided by the CSMC-RDouP is higher than the vanilla CSMC with a fixed computational budget, while the vanilla CSMC is better than CSMC-RDouP for trees with a small number of taxa.

\begin{figure}[htp]
\centering
\includegraphics[width=0.7\textwidth]{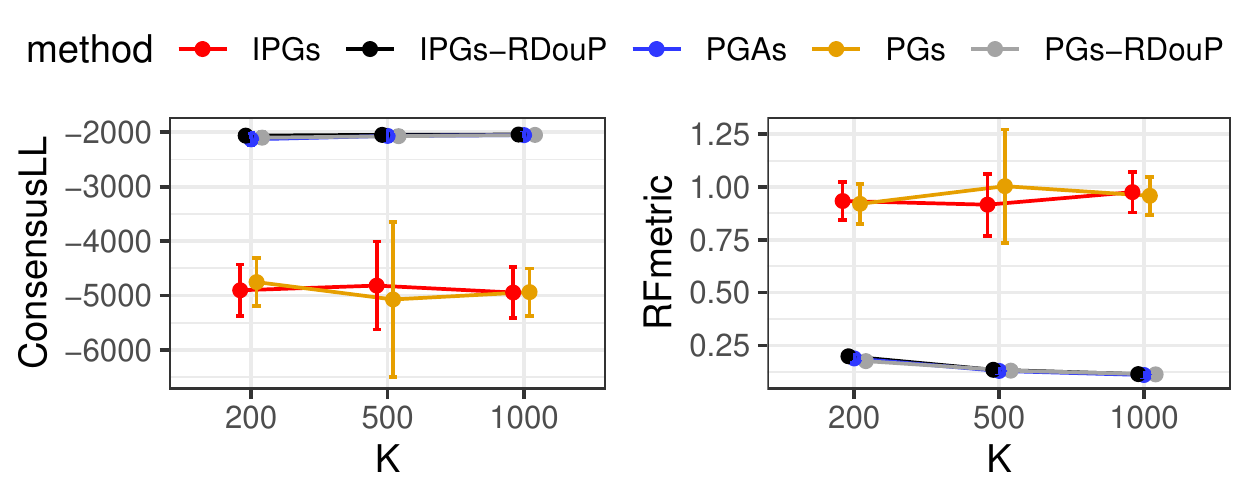}
\caption{Comparison of PGs and IPMCMC with the vanilla  CSMC (IPGs, PGs) and CSMC-RDouP  (IPGs-RDouP, PGs-RDouP), PGAs with CSMC-RDouP  (PGAs) as a function of number of particles. The x-axis represents number of particles. The three levels from left to right are $200$, $500$, $1000$,  respectively. The y-axis of represents the log likelihood of the consensus tree  and Robinson Foulds metric. 
} 
\label{fig:PGS-SM}
\end{figure}

We conducted another experiment to investigate the performance of  the vanilla CSMC and CSMC-RDouP  in PGs and IPMCMC,  as a function of the number of particles. We also investigated the performance of PGAs with CSMC-RDouP. 
Figure \ref{fig:PGS-SM} displays the comparison of PGs and IPMCMC with vanilla version of  CSMC (IPGs, PGs) and CSMC-RDouP  (IPGs-RDouP, PGs-RDouP) as a function of number of particles. 
For both PGs and IPMCMC with the vanilla CSMC, the log-likelihood of majority-rule  consensus tree and RF metric  do not improve if we increase $K$. PGs, IPMCMC, and PGAs with CSMC-RDouP  perform better in terms of the log-likelihood and RF metric. If we increased $K$, the log-likelihoods increase and RF metrics decrease. The log-likelihood and RF metric provided by PGs, IPMCMC, and PGAs with CSMC-RDouP  are close.

{\color{black} To understand the poor performance of the vanilla  CSMC in the particle Gibbs samplers, we investigate the ESS in the conditional CSMC algorithm, which is the main component of particle Gibbs. 
Table \ref{tab:ess} lists the mean ESS (0.025-quantile, 0.975-quantile)  from all the iterations of the conditional CSMC algorithms using 100 runs of PGs, PGs-RDouP, PGAs-RDouP, respectively, for 3 scenarios.  Although the mean ESS is generally larger in the vanilla  conditional CSMC than that in the conditional CSMC-RDouP,  there is a large percentage of cases in which the ESS is exactly equal to 1 (see the line with `ESS=1' in Table  \ref{tab:ess}), even when the number of taxa is as small as 7.  Consequently, the Markov chain will get stuck at the same tree topology as the reference trajectory and hardly move. In contrast, all the ESS values are larger than 1 in both PGs-RDouP and PGAs-RDouP. The Markov chains are able to explore different tree topologies rather than the one on the reference trajectory.

\begin{table}[h!]
  \caption{ESS (0.025-quantile, 0.975-quantile) in the conditional CSMC algorithms of PGs, PGs-RDouP, and PGAs-RDouP.}
    \label{tab:ess}
\begin{center}
{\small
 \begin{tabular}{c |c |c |c } 
 $\#$Taxa &$7$ &$10$  & $10$ \\  \hline
 Sequence length &$500$ &$2000$  & $2000$ \\ 
 \hline
$K$ &$20000$ &$10000$  & $20000$ \\ \hline
PGs &  128 (1.000, 247) &   21 (1.000, 85) &  63 (1.000, 208) \\ 
 ESS=1 & 0.6\% &33.3\%& 11.1\% \\ \hline 
PGs-RDouP  & 112 (1.094, 520) &  18 (1.011, 110) &   42 (1.014, 295)  \\ \hline
PGAs-RDouP &  112 (1.953, 517)  & 19 (1.014, 110)  & 41 (1.043, 300)    \\ 
\end{tabular}
}
\end{center}
\end{table}

We also find that,  with a fixed computational cost, high values of $K$ are more important than the number of PGs iterations.  In addition,  the computing speed of CSMC-RDouP can increase notably from parallelization. See the \emph{Supplementary Material Section 6} for details.
}

\section{Real data analysis}
\label{sec:preal}
We analyze two real datasets of DNA sequences. We assume the K2P model for evolutionary process, and make the clock assumption for trees ($t$).  
We consider the inference of $t$ and $\theta$ via particle Gibbs sampler, and evaluate the tree construction quality using the log-likelihood function of majority-rule  consensus tree.

The first real dataset we analyze is a set of DNA sequences for nine primates \citep{brown1982}.  In each DNA sequence, there are $888$ sites. As we have investigated in \emph{Simulation Studies}, with a fixed computational budget, the number of particles ($K$) is more important than the number of MCMC iterations ($N$) in improving the mixing of algorithm. We fix the number of MCMC iterations $N = 5000$, and vary $K$ to investigate the estimation by IPGs-RDouP and PGs. Table \ref{tab:prime} displays the log-likelihood of the consensus tree provided by IPGs-RDouP and PGs with different number of particles. We select four levels of $K$ for PGs, $K = 1000, 2000, 5000, 10000$. 
We set the total number of nodes for running conditional CSMC and CSMC algorithms to be twice as the number of nodes running conditional CSMC ($M = 2P$), and set $M = 4$. We set $K$ for PGs $4$ times as large as each worker of IPMCMC, to guarantee fixed computational budgets for the two methods. 
For each algorithm, we repeat  $10$ times. 

 Table \ref{tab:prime} displays the log-likelihood (mean and standard deviation) of the consensus tree obtained from PGs and PGs-RDouP, with different numbers of particles; each case is repeated  $10$ times with different initialization of evolutionary parameters and trees. {\color{black} The mixing of PG chains is poor even with $K = 10000$. Multiple chains with different initializations do not converge to the same posterior distribution. }The mixing of PGs-RDouP is improving when we increase the value of $K$. The log-likelihood gets higher and the standard deviation of log-likelihood decreases when we increase $K$. {\color{black} The  PGs-RDouP chain mixes well when $K = 10000$. Multiple chains with different initializations converge to the same  posterior distribution.} The mean and standard deviation for the posterior mean of $\theta$ for $10$ replications provided by PGs-RDouP are $3.68$ and  $0.0094$, respectively. For one run of IPGs-RDouP, 
 the posterior mean and $95\%$ credible interval of $\theta$ are $3.69$ and $(3.23, 4.23)$ respectively.

In addition, we run PGs-RDouP and PGAs using $K = 10000$ and $N = 5000$. {\color{black} For comparison, we also run MrBayes for $5\cdot 10^7$ iterations. The log-likelihood of the consensus tree provided by MrBayes is $-5601.4$. The majority-rule consensus tree provided by IPGs-RDouP, PGs-RDouP, PGAs and MrBayes are the same, as displayed in Figure 6 of \emph{Supplementary Material Section 7}.}

{\color{black} The second real data analysis is presented in \emph{Supplementary Material Section 7}.}

\begin{table}[h!]
  \caption{Log-likelihood of consensus trees provided by PGs and IPGs-RDouP, with mean (standard deviation), with varying numbers of particles.}
    \label{tab:prime}
\begin{center}
{\small
 \begin{tabular}{c |c |c |c |c} 
$K$ &$1000$ &$2000$  & $5000$ &$10000$\\ 
 \hline
PGs & -8424.1  (594.5)  &  -8247.0 (723.2)  &  -8333.8 (789.5)  & -7930.9 (549.9) \\ \hline
IPGs-RDouP & -5626.8  (1.9)  &  -5611.4 (1.7)  &  -5580.6 (1.3)  & -5553.8 (1.3) 
\end{tabular} 
}
\end{center}
\end{table}


\section{Conclusion}
\label{sec:pcon}

We have proposed a combinatorial sequential Monte Carlo method with an RDouP proposal.  Instead of randomly choosing a pair of trees to combine, we first use a revert step to find the reverted state of the current state, then in each merge step we randomly choose a pair of trees to combine. This proposal can benefit the exploration of tree posterior distribution. Our experimental results indicate that the RDouP proposal can improve the performance of CSMC, and this improvement can be enlarged when the number of taxa increases. The framework of CSMC-RDouP  is also easy to parallelize.  
This makes the proposed CSMC more scalable to large DNA datasets compared with traditional Bayesian methods, such as MCMC.

Since SMC is a non-iterative algorithm, particles that fail to survive in the current iteration will not have a chance to come back to be part of the future enlarged particles. Consequently, path degeneracy is inevitable for SMC algorithms.  This issue becomes more serious in conditional SMC algorithms. Therefore, methods that can mitigate the path degeneracy issue are crucial for improving the performance of SMC  and the corresponding particle Gibbs samplers. Our proposed novel RDouP proposal provides a simple but effective way to reduce the path degeneracy issue for the combinatorial SMC algorithms. The idea is to allow one chance to regret  the particle propagation  by undoing the last propagation and then redoing
the propagation twice.  Note that although our method was motivated and illustrated using the phylogenetic applications, it can also be applied to other cases in which the CSMC is used. Also, the RDouP proposal can be based on other proposal distributions besides the simple merge proposal.

We have presented a particle Gibbs sampler, a hybrid of  CSMC-RDouP and Gibbs sampler to estimate evolutionary parameters jointly with the phylogenetic trees. We have demonstrated the path degeneracy issue in the vanilla version of CSMC can be greatly mitigated in CSMC-RDouP. Consequently,  CSMC-RDouP can be used in the particle Gibbs for phylogenetics, which previously suffers from the problem of poor mixing of the Markov chain. Moreover,  CSMC-RDouP  can be used in more advanced particle Gibbs samplers, including but not limited to particle Gibbs with ancestor sampling and interacting particle Markov chain Monte Carlo to achieve further improvements. 

{\color{black} The consistency property of CSMC-RDouP holds when number of particles $K$ goes to infinity. However, $K$ cannot be made arbitrarily large  in practice  due to the memory limit. In addition,  the computational costs  of CSMC-RDouP increases linearly with number of particles $K$. A small value of $K$ may induce large bias for SMC estimates. 
Hence, it is important to select a proper value of $K$. \cite{doucet2015efficient} suggest selecting $K$ that the standard deviation of the log-likelihood estimate is around one so that the aysmptotic variance of the resulting PMCMC estimates is minimized with a fixed computational cost. }

There are several possible directions to refine this methodology.  First, we could explore more ways to allow the particles to undo  the previous  particle propagation.  For example, we can use a different base proposal distribution, and we can revert the particles by undoing two steps of the base propagation.  A second direction is to incorporate MCMC moves in CSMC-RDouP in a way that is described in \cite{Doucet11tutorial} to further mitigate the path degeneracy issue by jittering the particles.
Asymptotic variance of the SMC estimator was studied in \cite{chopin2004central}  under different resampling schemes, and displayed that advanced resampling schemes can reduce the variance of estimators. 
Another line of future work is to propose a computationally efficient  resampling scheme, which is both unbiased and admits lower asymptotic variance for discrete tree spaces \citep{fearnhead2003JRSS}.


\end{document}


\if0\blind
{
  \title{\bf Supplementary Material for the Bioinformatics submission ``Particle Gibbs Sampling for Bayesian Phylogenetic inference''}
\author{Shijia Wang
   \hspace{.2cm}\\
   School of Statistics and Data Science, LPMC \& KLMDASR, Nankai University, China\\
    Liangliang Wang \\
   Department of Statistics and Actuarial Science, Simon Fraser University, BC, Canada}  \maketitle
} \fi

\if1\blind
{
  \bigskip
  \bigskip
  \bigskip
  \begin{center}
    {\LARGE\bf Title}
\end{center}
  \medskip
} \fi

\section{K2P model}

The evolutionary model considered in this article is the K2P model. The rate matrix of the CTMC for K2P model is 
\[
Q = 
 \begin{pmatrix}
  - & \pi_{C} & \kappa\pi_{G} & \pi_{T} \\
  \pi_{A} & - & \pi_{G} & \kappa\pi_{T} \\
  \kappa\pi_{A}  & \pi_{C}  &- & \pi_{T}  \\
  \pi_{A} & \kappa\pi_{C} & \pi_{C} & - 
 \end{pmatrix}.
\]
Here the stationary state frequencies 
 $\pi_A=\pi_C=\pi_G=\pi_T=0.25$, and $\kappa$ represents the ratio of transition to transversion, which is the only parameter in the K2P evolutionary model ($\theta = \kappa$).

%

%

\section{Markov chain Monte Carlo for Bayesian phylogenetics}
Markov chain Monte Carlo (MCMC) algorithms are typically used to approximate $\pi(t, \theta)$ when we are able to evaluate $\gamma (t, \theta)$ pointwise. Under weak conditions, MCMC is guaranteed to converge to the posterior distribution asymptotically.
We sample from $\pi(t|\theta)$ and $\pi(\theta|t)$ iteratively until convergence is achieved.
 
Since $\pi(\theta|t)$ does not admit a closed form, we use one Metropolis-Hastings (MH) step to sample $\theta$. We let $q(\cdot|\theta)$ denote the proposal distribution given $\theta$, and sample $\theta^*\sim q(\cdot|\theta)$. 
 The {MH} ratio for accepting a newly proposed
$\theta^*$ is 
\begin{eqnarray}\label{eqn:mhRatioForTheta}
\alpha(\theta\to \theta^*|t) &=&\min\left(1,
\frac{p(y|{\theta^*},t)p(\theta^*) } {p(y|{\theta},t)p(\theta) }
\frac{q(\theta|\theta^*)}{q(\theta^*|\theta)}
\right).
\end{eqnarray}
Algorithm \ref{algo:MHwithinPGS} provides a detailed description of the MH algorithm for $\theta$. 
\begin{algorithm}
\caption{\bf{The accept-reject Metropolis Hasting algorithm for $\theta$}} 
\label{algo:MHwithinPGS}
\begin{algorithmic}[1]
  \State {\bfseries Inputs:} (a) Prior over evolutionary parameter $p(\theta)$;  
(b) Likelihood function $p(y | t, \theta)$; (c) One  phylogenetic tree sample $t$; (d) MH proposal $q(\cdot|\theta)$.

   \State {\bfseries Outputs:}  $\theta^*\sim \pi(\theta|t)$.



\State Propose a new evolutionary parameter, $\theta^*\sim q( \cdot|\theta)$.

\State 	Compute the Metropolis-Hastings ratio $\alpha(\theta\to \theta^*|t)$ according to Eq. (\ref{eqn:mhRatioForTheta}).
    
   \State Simulate $u\sim U(0,1)$.
    
\If{$u < \alpha(\theta\to \theta^*|t)$}
\State  $\theta^* = \theta^*$.   

  \Else
    \State $\theta^* = \theta$.
    
  \EndIf

\end{algorithmic}
\end{algorithm}

We also use MH algorithm to sample new trees $t$. The MCMC proposals for $t$ involve, for example, the multiplicative branch proposal \citep{Lakner01022008},  the stochastic nearest neighbor interchange (NNI) proposal \citep{Jow2002} and subtree prune and regraft (SPR) move \citep{allen2001subtree}.  Given evolutionary parameter $\theta$, the {MH} ratio for accepting a newly proposed $t^*$ is 
\begin{eqnarray}\label{eqn:mhRatioFort}
\alpha(t\to t^*|\theta) &=&\min\left(1,
\frac{p(y|{\theta},t^*)p(t^*) } {p(y|{\theta},t)p(t) }
\frac{q(t|t^*)}{q(t^*|t)}
\right).
\end{eqnarray}
In this article, we consider the K2P nucleotide substitution model, in which the only unknown parameter is $\kappa$.  A commonly used prior for parameter $\kappa$ is 
$\kappa \sim   \mbox{Exp}(\mu_0)$, and the proposal 
$q(\kappa^*|\kappa)=m \kappa$. Here $\mu_0$ is a hyper-parameter in exponential prior distribution, and $m \sim \mbox{Unif}(\frac{1}{a}, a)$ with a
prefixed tuning parameter $a>0$. The acceptance probability $\alpha$ for MH algorithm can be simplified as    
\begin{eqnarray*}
\alpha(\kappa\to \kappa^*|t) =\min\left(1,
\frac{p(y|{\kappa^*},t)} {p(y|{\kappa},t)}\cdot
e^{\mu_0\kappa(1-m)}\cdot m  \right).
\end{eqnarray*}

\section{CSMC with the RDouP proposal}

To facilitate the discussion of $m^{+}$, $m^{-}$, and $r^{-}$ for the phylogenetic tree application, we rewrite Equation (9) in \emph{Section 3.2 of the Main Manuscript} as follows, 
\begin{align} \nonumber \label{eqn: base_proposal1}
&w(\tilde \aus_{r-1,k}, \aus_{r,k}) \\ 
= &  \frac{\gamma(\sls(\tilde \aus_{r-1,k})}{\gamma(\varrho(\tilde \aus_{r-1,k}))} \cdot  \frac{m_{\sls(\tilde \aus_{r-1,k})}^{-}(\varrho(\tilde \aus_{r-1,k}))}{m_{{\varrho}(\tilde \aus_{r-1,k})}^{+}( \sls(\tilde \aus_{r-1,k}))} \\
  & \cdot \frac{\gamma({\color{black}\beta (\aus_{r,k}))}}{\gamma(\sls(\tilde \aus_{r-1,k})} \cdot  \frac{m_{\aus_{r,k}}^{-}(\sls(\tilde \aus_{r-1,k}))}{m_{\sls(\tilde \aus_{r-1,k})}^{+}(\aus_{r,k})}\\  \label{eqn: base_proposal3}
  & \left.  \middle/\frac{\gamma({\color{black}\beta(\tilde \aus_{r-1,k}))}}{\gamma(\varrho(\tilde \aus_{r-1,k}))} \cdot  \frac{1}{r_{\varrho(\tilde \aus_{r-1,k})}^{-}(\tilde \aus_{r-1,k})}) \right..
\end{align}

We let {\color{black} $r^{-}_{\varrho(\aus)}(\aus) = \check{m}^{+}_{\varrho(\aus)}(\aus)$, where $\check{m}^{+}_{\varrho(\aus)}(\aus) = m^{+}_{\varrho(\aus)}(\beta(\aus))$.}
The common component of Equations (\ref{eqn: base_proposal1} - \ref{eqn: base_proposal3})  is 
\begin{align} \label{eqn: mainweightupdate}
u(s, s')  = \frac{\gamma(s')}{\gamma(s)} \cdot  \frac{1}{m_{s}^{+}(s')},
\end{align}
where {\color{black} $s, s' \in \poset$}, and the rank of $s'$ is 1 higher than the rank of $s$.

We assume that the subtrees being merged in $s$ are $(t_{1}, X_{1})$ and $(t_{2}, X_{2})$, and that the merged subtree connecting $(t_{1}, X_{1})$ and $(t_{2}, X_{2})$ is $(t_{m}, X_{m})$. In other words, 
$s' =  s 
 \cup \{(t_\textrm{m}, X_\textrm{m})\} \backslash \{(t_1, X_1), (t_2, X_2)\}. $
 
We consider the clock trees and non-clock trees respectively. For clock trees, the increment of the height of the forest $(\Delta_{h})$ is proposed from an exponential distribution with rate $\lambda_s$ (the prior distribution). 
The density of the proposal $m^{+}$ is 
\[m_s^{+}(s') = \binom{|s|}{2}^{-1}\lambda_s\cdot\exp(-\lambda_s\Delta_{h}),\] 
where $\lambda_s$ denotes the rate of the exponential distribution that proposed the incremental height of forests $(\Delta_{h})$. 
The proposal in  non-clock tree case is similar to the clock case except for the branch length. For the proposal over branch lengths in non-clock tree, we consider two cases based on the number of trees in the forest. If there
are exactly two trees in the forest before the propagation, we propose
a single length $(b_{1})$ distributed according to an exponential distribution with rate $\lambda$ (the prior distribution). Otherwise if there
are more than two trees, we propose two independent branch lengths $b_{1}$, $b_{2}$ according to an exponential distribution with rate $\lambda$ (the prior distribution).  The probability density of the base proposal $m^{+}$ is 
\[
 m^{+}_{s}(s') =\binom{|s|}{2}^{-1}\lambda\cdot\exp(-\lambda b_{1})\times \bigg(\boldsymbol{1}[|s| = 2]+\lambda \exp(-\lambda b_{2})\boldsymbol{1}[|s| > 2] \bigg).
\]
With the above choice of the base proposals,  Equation (\ref{eqn: mainweightupdate}) can be simplified to 
\begin{eqnarray*} 
u(s, s')   = \frac{p(y(X_\textrm{m})|\theta,t_\textrm{m})}{p(y(X_\textrm{1})|\theta,t_1)p(y(X_\textrm{2})|\theta,t_2)}.
\end{eqnarray*}

For clock trees, the probability of the backward base kernel $m^{-}$ is 1;  and for non-clock trees,
\[
 m^{-}_{s'}(s) = \frac{1}{\sum_{(t_i, X_i)\in s} I[|X_i| > 1]},
\] 
where $I[|X_i| > 1]$ denotes the number of non-trivial trees in forest $t_{i}$.

In summary, for clock trees,
\begin{align*} 
&w(\tilde \aus_{r-1,k}, \aus_{r,k}) \\
= & u(\varrho(\tilde \aus_{r-1,k}), \sls(\tilde \aus_{r-1,k})) u(\sls(\tilde \aus_{r-1,k}), {\color{black}\beta(\aus_{r,k}))}/ u(\varrho(\tilde \aus_{r-1,k}), {\color{black}\beta(\tilde \aus_{r-1,k})});
\end{align*}
and for non-clock trees, 
\begin{align*} 
&w(\tilde \aus_{r-1,k}, \aus_{r,k})  \\ 
=& \left[u(\varrho(\tilde \aus_{r-1,k}), \sls(\tilde \aus_{r-1,k})) u(\sls(\tilde \aus_{r-1,k}), {\color{black}\beta(\aus_{r,k}))}/ u(\varrho(\tilde \aus_{r-1,k}), {\color{black}\beta(\tilde \aus_{r-1,k})}) \right]\\
& \cdot \frac{1}{\sum_{(t_i, X_i)\in \varrho(\tilde \aus_{r-1,k})} I[|X_i| > 1]} \cdot \frac{1}{\sum_{(t_i, X_i)\in \sls(\tilde \aus_{r-1,k})} I[|X_i| > 1]}.
\end{align*}

\section{Algorithmic descriptions}


Algorithm \ref{algo:condition-smc} describes the conditional CSMC-RDouP algorithm.   
Algorithm  \ref{algo:pmcmc-PGAS} describes the conditional CSMC-RDouP with ancestor sampling. 
Algorithm  \ref{algo:pmcmc-PGS} shows the particle Gibbs sampler. 

\begin{algorithm}
\caption{: \textbf{Conditional CSMC-RDouP algorithm}} 
\label{algo:condition-smc}
\begin{algorithmic}[1]
  \State {\bfseries Inputs:} (a) Prior over {\color{black} augmented} partial states $p(s)$;  
(b) Likelihood function $p(y | s, \theta)$; (c) One pre-specified particle trajectory for partial states $s^*_{1:R}$. 

   \State {\bfseries Outputs:}  Approximation of the posterior distribution, $\sum_k W_{R,k} \delta_{ s_{R,k}}(\cdot) \approx \pi(\cdot)$.
	\State  Initialize SMC iteration index: $r \leftarrow 1$.  

\For{ $r \in\{1, \dots, R\}$}
 \State Initialize $s_{1, r} = s^*_{r}$.
 \EndFor

 \For{$k\in\{1, \dots, K\}$}
  \State Initialize particles with the least partial state $\perp$. 
  \State  Initialize weights: $w_{1,k} \leftarrow 1$; $W_{1,k} \leftarrow 1/K$.
  \EndFor

\For{ $r \in\{2, \dots, R\}$}
\If{$\text{rESS}( W_{r-1,\cdot}) < \epsilon$ }

\State Resample the particles with index $k = 2, 3, \ldots, K$. Note that we do not resample particles in the reference trajectory. 
   \For{$k \in \{1, \dots, K\}$}
    \State \label{step:adapt-reset-weights} Reset particle weights: $\tilde w_{r-1,k} = 1$; $\tilde W_{r-1,k} = 1/K$.
    \EndFor
    \Else
    \For{$k \in \{1, \dots, K\}$}
    \State $\tilde w_{r-1,k} = w_{r-1,k} ;\tilde s_{r-1,k} =  s_{r-1,k} $. 
    \EndFor
\EndIf
 \For{  $k\in\{1, \dots, K\}$}
      \If{$k > 1$ }
       \State Sample particles $ s_{r,k} \sim \nu^+_{ \tilde s_{r-1,k}}(\cdot)$, using one revert move to find $\varrho(\tilde s_{r-1,k})$, and two base proposals to propose $\sls(\tilde s_{r-1,k})$ and $s_{r,k}$. 
       \EndIf
         \State Compute pre-resampling normalized weights: $$ w_{r,k} =  \tilde w_{r-1,k}\cdot \frac{\gamma(s_{r,k})\cdot \nu_{s_{r,k} }^-( \tilde s_{r-1,k})}{\gamma(\tilde s_{r-1,k})\cdot\nu_{\tilde s_{r-1,k}}^+(s_{r,k})}.$$

  \EndFor
\State Construct $$\pi_{r,K}(s) = \frac{1}{K} \sum_{k=1}^K W_{r,k} \delta_{s_{r,k}}(s).$$
  \EndFor
\end{algorithmic} 
 \end{algorithm}

\begin{algorithm}
\caption{: \bf{Conditional CSMC-RDouP with ancestor sampling}} 
\label{algo:pmcmc-PGAS}

\begin{algorithmic}[1]
  \State {\bfseries Inputs:} (a) Prior over {\color{black} augmented} partial states $p(s)$;  
(b) Likelihood function $p(y | s, \theta)$; (c) One pre-specified particle trajectory for {\color{black} augmented} partial states $s^*_{1:R}$. 

   \State {\bfseries Outputs:}  Approximation of the posterior distribution, $\sum_k W_{R,k} \delta_{ s_{R,k}}(\cdot) \approx \pi(\cdot)$.
	\State  Initialize SMC iteration index: $r \leftarrow 1$.  

\For{ $r \in\{1, \dots, R\}$}
 \State Initialize $s_{1, r} = s^*_{r}$.
 \EndFor

 \For{$k\in\{1, \dots, K\}$}
  \State Initialize particles with the least partial state $\perp$. 
  \State  Initialize weights: $w_{1,k} \leftarrow 1$; $W_{1,k} \leftarrow 1/K$.
  \EndFor

\For{ $r \in\{2, \dots, R\}$}
\If{$\text{rESS}( W_{r-1,\cdot}) < \epsilon$}

\State Resample the particles with index $k = 2, 3, \cdots, K$. Note that we do not resample particles in the reference trajectory. 
   \For{$k \in \{1, \dots, K\}$}
    \State \label{step:adapt-reset-weights} Reset particle weights: $\tilde w_{r-1,k} = 1$; $\tilde W_{r-1,k} = 1/K$.
    \EndFor
    \Else
    \For{$k \in \{1, \dots, K\}$}
    \State $\tilde w_{r-1,k} = w_{r-1,k} ;\tilde s_{r-1,k} =  s_{r-1,k} $. 
    \EndFor
\EndIf
\If{$r \geq 3$}
  \For{$k \in \{1, \dots, K\}$}
  \State Compute the probability of connecting the partial reference trajectory $s_{ r:R, 1}$ to one of the history trajectories $s_{1:r-1,k}$,
  \[
P(a_{r} = k)\propto w_{ r-1, k}\cdot\frac{\gamma(s_{r-1, k}, s_{r, 1})}{\gamma(s_{ r-1, k})}. 
\]

   \EndFor
\State  We sample ancestor index $a_{r} = k$ with probability $P(a_{r} = k)$.

\EndIf
 \For{  $k\in\{1, \dots, K\}$}
      \If{$k > 1$ }
       \State Sample particles $ s_{r,k} \sim \nu^+_{ \tilde s_{r-1,k}}(\cdot)$, using one revert move to find $\varrho(\tilde s_{r-1,k})$, and two base proposals to propose $\sls(\tilde s_{r-1,k})$ and $s_{r,k}$. 
       \EndIf
         \State Compute pre-resampling normalized weights: $$ w_{r,k} =  \tilde w_{r-1,k}\cdot \frac{\gamma(s_{r,k})\cdot \nu_{s_{r,k} }^-( \tilde s_{r-1,k})}{\gamma(\tilde s_{r-1,k})\cdot\nu_{\tilde s_{r-1,k}}^+(s_{r,k})}.$$

  \EndFor
\State Construct $$\pi_{r,K}(s) = \frac{1}{K} \sum_{k=1}^K W_{r,k} \delta_{s_{r,k}}(s).$$
  \EndFor
\end{algorithmic} 
 \end{algorithm}

\begin{algorithm}
\caption{: \bf{Particle Gibbs sampler}} 
\label{algo:pmcmc-PGS}
\begin{algorithmic}[1]
  \State {\bfseries Inputs:} (a) Prior over {\color{black} augmented} partial states $p(s)$ and evolutionary parameter $p(\theta)$;  
(b) Likelihood function $p(y | s, \theta)$; (c) One pre-specified particle trajectory for {\color{black} augmented} partial states $s^*_{1:R}$; (d) MH proposal for $\theta$, $q(\cdot|\theta)$.

   \State {\bfseries Outputs:}  Approximated posterior distribution $\hat{\pi}(\theta, t)$.

\State Initialize MCMC iteration index $i\leftarrow 0 $.
\State Initialize $\theta(0)$ to an arbitrary value. 
\State Initialize the reference trajectory of the conditional CSMC-RDouP (Algorithm \ref{algo:condition-smc}) to $s^*_{1:R}$. (Without generality we could set $s_{1:R, 1}(0) = s^*_{1:R}$.)
\State Run the conditional CSMC-RDouP algorithm  targeting $\pi(t|\theta(0))$, sample 
$s_{1:R}^*\sim \hat{\pi}(\cdot|\theta(0))$.
\For{$i =  1, 2, \ldots, N$}
\State Sample $\theta(i)\sim p(\theta|y, t = s_{R}^*)$ using the accept-reject Metropolis Hastings algorithm.
\State Update the reference trajectory to $s^*_{1:R}$ and run the conditional CSMC-RDouP algorithm (Algorithm \ref{algo:condition-smc}) targeting $\pi(t|\theta(i))$, sample 
$s_{1:R}^*\sim \hat{\pi}(\cdot|\theta(i))$.
\EndFor
\State Burn in a proportion (e.g. $50\%$) of the PGs chain. 
\end{algorithmic}
\end{algorithm}

\section{{\color{black} PGAs and IPMCMC}}
{\color{black} \subsection{Implementation of particle Gibbs sampler with ancestor Sampling}
We turn to the implementation of the ancestor sampling step. Without loss of generality, we let $\aus_{1:R, 1}$ be the reference trajectory. For rank $r\geq 3$, we consider the part of the reference trajectory $\aus_{r:R, 1}$ 
ranging from the current partial state $r$ to the final state $R$. The objective is to connect the partial reference trajectory $\aus_{ r:R, 1}$ to one of the historical trajectories $\{\aus_{1:r-1,i}\}_{i = 1}^{K}$. We use notation $a_{r}\in\{1,2,\ldots, K \}$ to denote the index of ancestor sampling trajectory for partial state of rank $r$.

We assign  each index $a_{r}\in\{1,2,\ldots, K \}$ a weight function, which admits the form
\[
P(a_{r} = k)\propto  w_{r-1,k}\cdot\frac{\gamma(\aus_{1:r-1,k}, \aus_{r:R, 1})}{\gamma(\aus_{1:r-1,k})} \propto w_{ r-1, k}\cdot\frac{\gamma(\aus_{r-1, k}, \aus_{r, 1})}{\gamma(\aus_{ r-1, k})}. 
\]
We sample $a_{r}$ with probability $P(a_{r} = k)$. The invariance property for the kernel of PGAs still holds. We described the PGAs in \emph{Supplementary Section 2}. The following assumption describes the  condition on $P(a_{r} = k)$.
 \setcounter{theorem}{1}
{\assumption For all $\aus_{r-1,k}$, $\aus_{r, 1}\in\poset$, $\nu^{+}_{\aus_{r-1, k}}(\aus_{r, 1}) = 0$ implies $P(a_{r} = k) = 0$.\\}
\noindent We use an example to illustrate the benefits of using the proposed  CSMC-RDouP  algorithm in the ancestor sampling step. Figure \ref{fig:PGAS} provides an example of the ancestor sampling step with CSMC-RDouP  and the vanilla CSMC, respectively.   To sample the ancestor of the reference trajectory at rank $r = 3$, we evaluate $P(a_{r} = k)$ for $k = 1, \ldots, K$. Using Assumption 2, $P(a_{3} = 1) = 0$ and $P(a_{3} = 3) = 0$ in PGAs with the vanilla  CSMC as $\nu^{+}_{\aus_{1,1}}(\aus_{2,2}) = 0$ and $\nu^{+}_{\aus_{1,3}}(\aus_{2,2}) = 0$. A new partial state can only be proposed by randomly choosing a pair of subtrees to combine. Hence, $\aus_{r}$ can connect to $\aus_{r-1}$ if and only if $\aus_{r-1}$ is the parent of $\aus_{r}$. As a result, the sampled ancestor is still the partial reference trajectory $\aus_{1:r-1}$. In contrast, if the RDouP proposal is used in PGAs, $P(a_{3} = 1) > 0$ and $P(a_{3} = 3) > 0$. 

\begin{figure}
\centering
\includegraphics[width=0.95\linewidth]{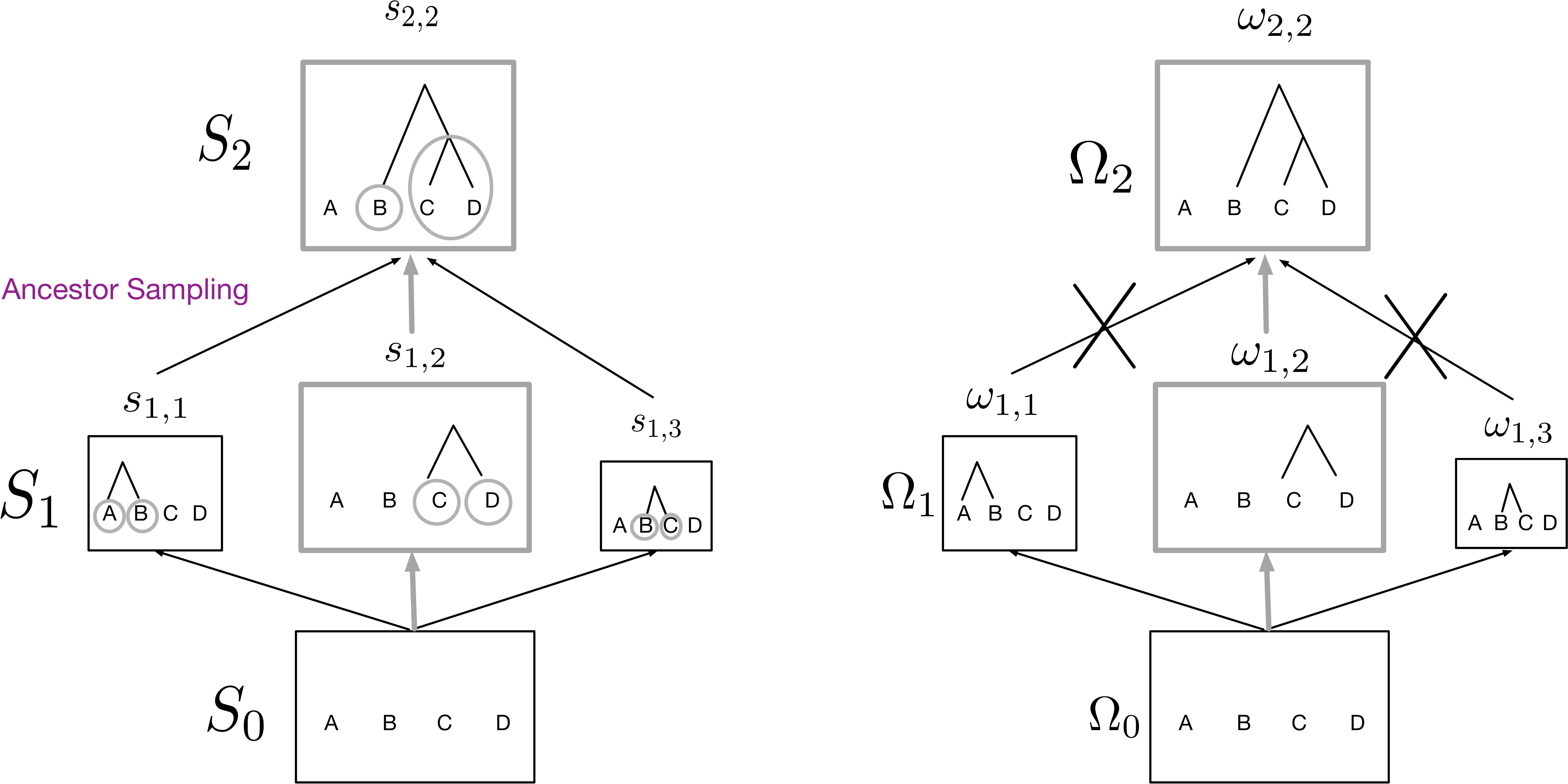}
\caption{An example of the ancestor sampling step with CSMC-RDouP (left) and the vanilla CSMC (right).  The bold grey trajectory represents the reference trajectory. The two grey circles of each particle indicate the two subtrees that are just merged.}
\label{fig:pgas}
\label{fig:PGAS}
\end{figure}
}

\subsection{Algorithmic description of Interacting Particle Markov chain Monte Carlo }
Algorithm \ref{algo:ipmcmc} summarizes the IPMCMC algorithm. For the sampling of $\theta$, we use the Metropolis-Hasting algorithm described in Algorithm \ref{algo:MHwithinPGS}. This step is same as PG. The updating of $t$ involves communication between several standard and conditional CSMC-RDouP algorithms. These algorithms are referred to as `nodes', and each of these nodes are assigned with index $m\in \{1, 2, \ldots, M\}$. 
Among these $M$ nodes, we run conditional CSMC-RDouP among $P$ of them, and the index of these conditional nodes are denoted as $c_{j}$, $j = 1, \ldots, P$. Among the rest $M-P$ nodes, we run standard CSMC-RDouP algorithm. A superscript is introduced to represent the index of nodes. For example, $s_{r,k}^{m}$ denotes the $k$th particle for the partial state of rank $r$ of the $m$th node. We introduce another notation $\hat{Z}$ to denote the marginalized likelihood estimates. 
The marginalized likelihood for node $m$ can be computed as
\[
\hat{Z}^{m} = \prod_{r = 2}^{R}\frac{1}{K}\sum_{k=1}^{K}w_{r,k}^{m}.
\]
After computing the marginalized likelihood for each node. We sample new index for each node $c_{j}$ with
\begin{eqnarray*}
P(c_{j} = m|c_{1:P\setminus j} ) &=& \hat{\xi}_{j}^{m},\\
\hat{\xi}_{j}^{m} &=& \frac{\hat{Z}^{m}\boldsymbol{1}_{m\notin c_{1:P\setminus j}}}{\sum_{n=1}^{M}\hat{Z}^{n}\boldsymbol{1}_{n\notin c_{1:P\setminus j}}}.
\end{eqnarray*} 
Then we sample the new reference trajectory for node $c_{j}$ with
\[
P(b_{j} = k|c_{j}) = w_{k, R}^{c_{j}}.
\]
The updated reference trajectory for node $c_{j}$ is $s_{1:R}^{c_{j},*} = s_{b_{j}, 1:R}^{c_{j}}$.
The switch of nodes is the essential part of IPMCMC.

\begin{algorithm}
\caption{: \bf{Interacting particle Markov chain Monte Carlo}} 
\label{algo:ipmcmc}
\begin{algorithmic}[1]
  \State {\bfseries Inputs:} (a) Prior over {\color{black} augmented} partial states $p(s)$ and evolutionary parameter $p(\theta)$;  
(b) Likelihood function $p(y | s, \theta)$; (c) Pre-specified particle trajectories for nodes; (d) MH proposal for $\theta$, $q(\cdot|\theta)$; (e) number of nodes $M$ and number of conditional nodes $P$. 

   \State {\bfseries Outputs:}  Approximated posterior distribution $\hat{p}(\theta, t|y)$.

\State Initialize MCMC iteration index $i\leftarrow 0 $.
\State Initialize $\theta(0)$ to an arbitrary value. 
\State Initialize an index $m = \{1, 2, \ldots, M\}$ to each node.
\For{$m \in  \{c_{1}, c_{2}, \ldots, c_{P}\}$}
\State Initialize the reference trajectory of each conditional node $m$ to $s^{m,*}_{1:R}$. 
\EndFor

\For{$m =  1, 2, \ldots, M$}
\For{$m \in  \{1:M\setminus c_{1:P}\}$}
\State Run the CSMC-RDouP algorithm  targeting $p(t|y,\theta(i-1))$.
\State Update the marginalized likelihood $\hat{Z}^{m} = \prod_{r = 2}^{R}\frac{1}{K}\sum_{k=1}^{K}w_{r,k}^{m}$.
\EndFor
\For{$m \in  \{c_{1}, c_{2}, \ldots, c_{P}\}$}
\State Run the conditional CSMC-RDouP algorithm  targeting $p(t|y,\theta(i-1))$, conditional on the reference trajectory $s_{1:R}^{c_{j},*}$.
\State Update the marginalized likelihood $\hat{Z}^{m} = \prod_{r = 2}^{R}\frac{1}{K}\sum_{k=1}^{K}w_{r,k}^{m}$.

\State Update index $c_{j}$ by sampling with probability
\begin{eqnarray*} 
P(c_{j} &=& m|c_{1:P\setminus j} ) = \hat{\xi}_{j}^{m},\\
\hat{\xi}_{j}^{m} &=& \frac{\hat{Z}^{m}\boldsymbol{1}_{m\notin c_{1:P\setminus j}}}{\sum_{n=1}^{M}\hat{Z}^{n}\boldsymbol{1}_{n\notin c_{1:P\setminus j}}}.
\end{eqnarray*} 
\State Update index $b_{j}$ with probability
\[
P(b_{j} = k|c_{j}) = w_{k, R}^{c_{j}}.
\]
\State Update the reference trajectory $s_{1:R}^{c_{j},*} = s_{b_{j}, 1:R}^{c_{j}}$.

\EndFor
\State Randomly sample a reference trajectory $s_{1:R}^{*}$ from $\{s_{1:R}^{c_{j},*}\}_{j = 1, \ldots, P}$.
\State Sample $\theta(i)\sim p(\theta|y, t = s_{R}^*)$ using the accept-reject Metropolis Hastings algorithm.
\EndFor
\State Burn in a proportion of the IPMCMC chain. \Comment{For example, we burn in $50\%$ of the chain in our experiment. }
\end{algorithmic}
\end{algorithm}

\section{Simulation Studies}
We evaluate the quality of tree reconstruction using the CSMC-RDouP  algorithm. 
We generate clock trees  assuming the branch length is exponentially distributed with rate $10\cdot \binom{|\aus_{r}|}{2}$, where $\aus_{r}$ denotes  a partial state of rank $r$.
Given the simulated tree, we simulate DNA sequences  using the {K2P} model with parameter $\theta=2.0$. 
Site independence is assumed. 
We use the \emph{majority-rule  consensus tree} to summarize the weighted samples of phylogenetic trees \citep{felsenstein1981evolutionary}.  We mainly use the
Robinson-Foulds (RF) metric based on sums of differences in branch lengths 
metric \citep{RobinsonFoulds1979}, and the partition metric ({PM}), also known as the symmetric difference or topology only RF metric  \citep{robinson1981} to measure the quality of the estimated tree. Stratified resampling \cite{Kitagawa1996} is conducted to resample particles when the rESS falls below $0.5$.

\subsection{Comparison of CSMC and CSMC-RDouP}
In this study, we first emphasize a comparison of the vanilla CSMC and the proposed CSMC-RDouP in terms of computational speed. We simulated $50$ clock trees, each of which has  $20$ leaves and $1000$ nucleotides on each leaf. We ran the two CSMC algorithms using $10000$ particles on all datasets. 
The relative runtime of CSMC-RDouP  compared with CSMC is about 1.4.  The computational speed of CSMC-RDouP  is lower than that of the vanilla  CSMC. This is expected as the proposal in the vanilla CSMC is simpler, while in our new proposal we have to use one move to find the reverted state and merge twice to propose the new partial state. The weight update function in CSMC-RDouP  algorithm is also more complicated.

\begin{figure}[htp]
\centering
\includegraphics[width=0.9\textwidth]{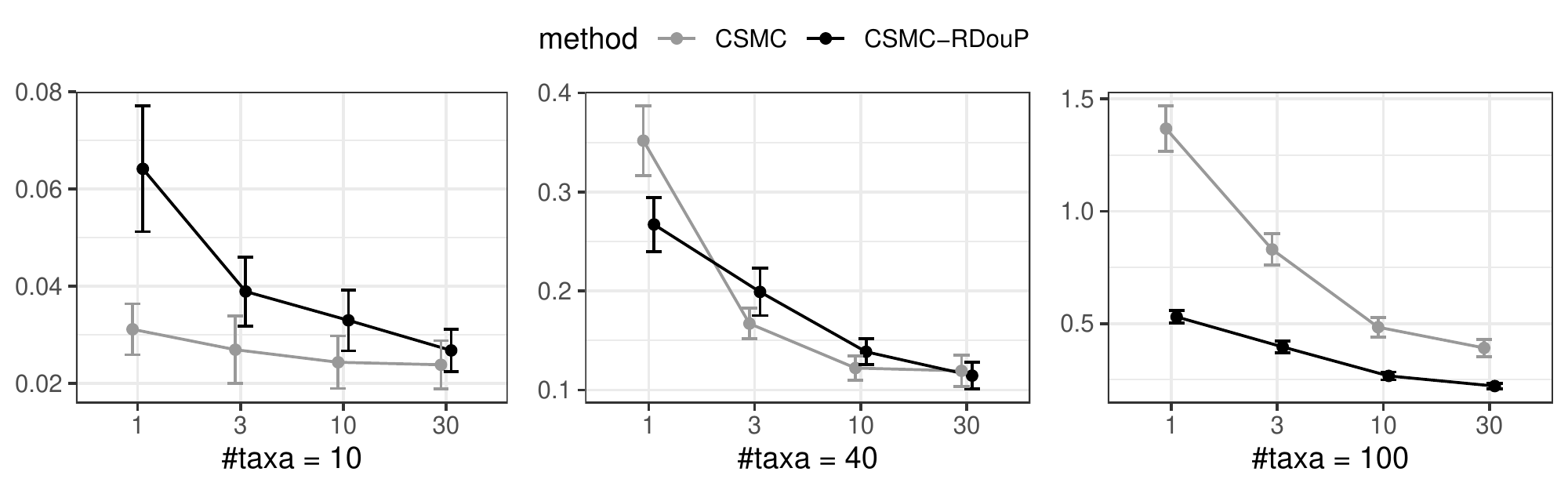}
\caption{{\color{black} Comparison of CSMC and CSMC-RDouP  as a function of number of (1000) particles in three scenarios: $10$ taxa (left), $40$ taxa (middle), $100$ taxa (right).  The x-axis represents the number of a thousand particles.  The four levels of $K$ (x-axis) from left to right of each panel for  CSMC-RDouP are $1\cdot 10^{3}$, $3\cdot 10^{3}$, $1\cdot 10^{4}$ and $3\cdot 10^{4}$ respectively. The four levels of $K$ (x-axis) from left to right of each panel for vanilla version of  CSMC are $1.5\cdot 10^{3}$, $4.5\cdot 10^{3}$,  $1.5\cdot 10^{4}$, $4.5\cdot 10^{4}$ respectively. The y-axis represents the Robinson Foulds metric. The RF metric decreases with incremental numbers of particles. For trees with a small number of taxa (e.g. 10), the tree reconstruction quality of the vanilla CSMC is better than CSMC-RDouP. For larger trees, the RF metric provided by the CSMC-RDouP is lower than the vanilla CSMC with a fixed computational budget. The gap in the RF metric between the two CSMCs widens as the number of taxa increases. }}
\label{fig:CSMC-RM}
\end{figure}

{\color{black} We did another experiment to evaluate the quality of tree reconstruction of the two CSMC algorithms. 
We simulated $60$ clock trees, with the number of leaves taken from the set $\{10, 20, 40\}$ ($20$ trees of each size). One dataset was simulated for each tree. The number of nucleotides on each leaf was $10,000$. We used the RF metric to measure the distance between the estimated consensus tree and the simulated reference tree. 
We compared two CSMC algorithms as a function of the number of particles. The number of particles for  CSMC-RDouP is $\{1\cdot 10^{3}, 3\cdot 10^{3}, 1\cdot 10^{4}, 3\cdot 10^{4}\}$. As we have mentioned in the previous experiment, the computational cost of CSMC-RDouP  is higher with a fixed number of particles. We selected four levels for the number of particles in the vanilla version of  CSMC,  $\{1.5\cdot 10^{3}, 4.5\cdot 10^{3}, 1.5\cdot 10^{4}, 4.5\cdot 10^{4}\}$, to guarantee the computational budget allocated for the two methods are close, with a slight favour for the vanilla CSMC. 
Figures \ref{fig:CSMC-RM} displays the RF metric of the CSMC-RDouP  and the vanilla CSMC algorithms as a function of the number of particles across different numbers of taxa. The RF metric decreases with increasing numbers of particles. 
For trees with a small number of taxa (e.g. $10$), the tree reconstruction quality of the vanilla CSMC is better than CSMC-RDouP. For larger trees, the RF metric provided by the CSMC-RDouP  is lower than the vanilla CSMC with a fixed computational budget.  The gap in the RF metric between the two CSMCs widens as the number of taxa increases. 
}

{\color{black} Finally, we illustrate the computational gains from parallelization, which is an advantage over standard MCMC approaches. We run CSMC-RDouP on one simulated data set from a clock tree, with $30$ taxa and $1000$ sites. We use $10,000$ particles and vary the number of threads on a 2.3 GHz Intel Core i9 processor. For each number of threads, we repeat CSMC-RDouP $50$ times. Figure \ref{fig:threads} shows the running time in milliseconds versus
different number of cores for the simulated data. The computing speed of CSMC-RDouP increases notably by adding cores.  We parallelize the propagation and weighting steps of CSMC-RDouP. 

\begin{figure}[htp]
\centering
\includegraphics[width=0.8\textwidth]{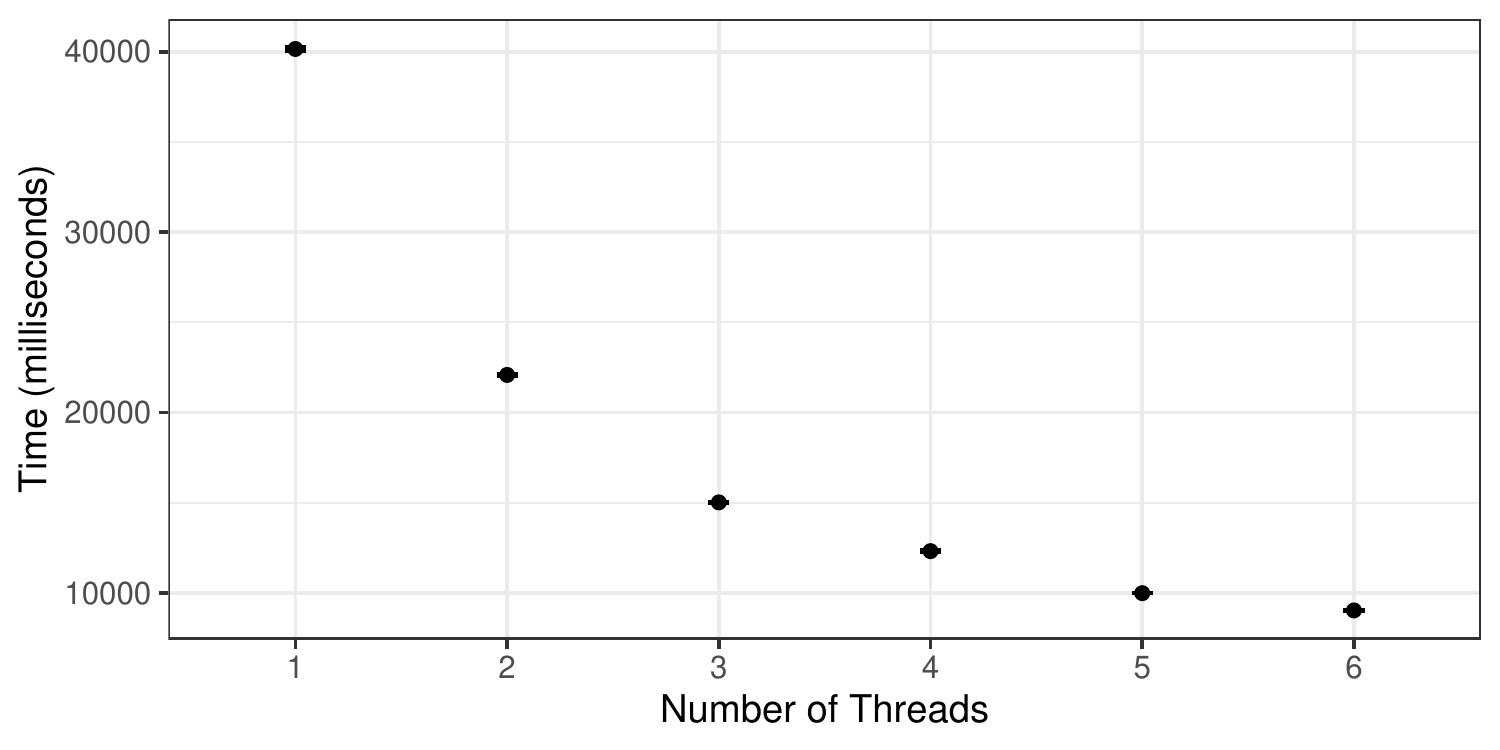}
\caption{{\color{black} Illustration of the parallelization of CSMC-RDouP.  The x-axis represents the number of threads. The y-axis represents the running time in milliseconds. The running time decreases with increasing numbers of threads. The mean running time for each value of threads (from $1$ to $6$) and the corresponding $95\%$ confidence intervals are $40152.68~ (40054.25, 40251.11)$,  $22082.36~(22021.31, 22143.41)$, $15018.82~(14974.65, 15062.99)$, $12316.84~(12270.06, 12363.62)$,  $9989.28~(9967.33, 10011.24)$,  $9029.08~(8998.91, 9059.26)$ respectively. }}
\label{fig:threads}
\end{figure}

}

\subsection{Comparison of particle Gibbs samplers}
In this section, we focus on joint estimation of evolutionary parameter $\theta$ and phylogenetic tree $t$ in the framework of particle Gibbs (PG) sampler.

\begin{figure}[htp]
\centering
\includegraphics[width=0.8\textwidth]{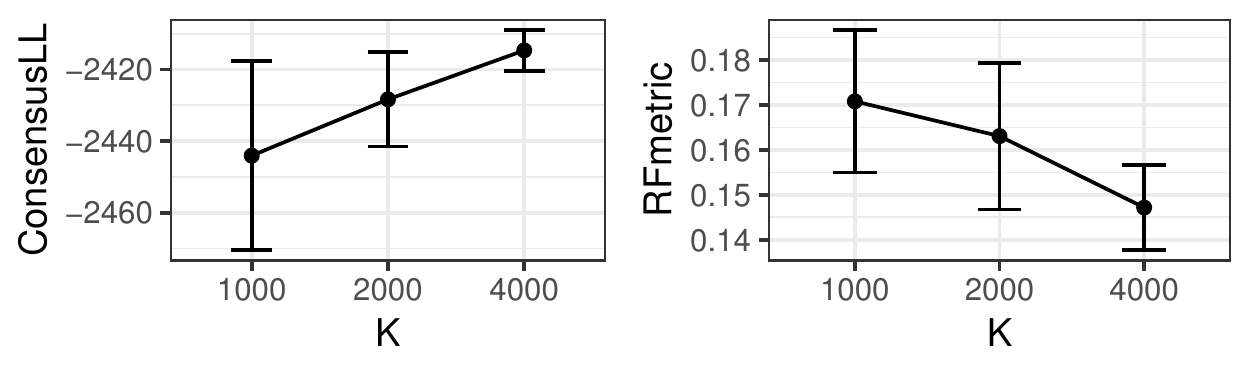}
\caption{Comparison of PGs with the RDouP proposal with  a fixed computational budget. {\color{black} The computational budgets are fixed at $K\cdot N = 2\cdot 10^{6}$, and three levels of $(K, N)$ are chosen: $(1000, 2000)$ (left), $(2000, 1000)$ (middle), $(4000, 500)$ (right). The PGs sampling with the RDouP proposal is repeated $10$ times for each scenario. With a higher value of $K$ (and lower number of MCMC iterations), the log-likelihood of \emph{majority-rule  consensus tree} increases and the RF metric decreases. This indicates that with a fixed computational cost, high values of $K$ are more important than the number of PGs iterations. 
}}
\label{fig:square}
\end{figure}

The computational cost of particle Gibbs samplers is a linear function of the number of particles $K$ and number of MCMC iterations $N$. Increasing both $K$ and $N$ can improve the mixing of particle Gibbs sampling. We first use one experiment to investigate the relative importance of $K$ and $N$ in improving the mixing of PGs with a fixed computational cost. We simulated $1$ clock tree with  $15$ leaves and one dataset with 500  nucleotides for each leaf. 
We evaluated the performance of the PGs  by the log-likelihood of \emph{majority-rule  consensus tree}  and the tree distance between the estimated consensus tree and the simulated reference tree  using the RF metric.  
We fixed the computational budgets at $K\cdot N = 2\cdot 10^{6}$, and chose three levels of $(K, N): (1000, 2000)$, $(2000, 1000)$, $(4000, 500)$. We repeated the PGs sampling with the RDouP proposal $10$ times. Figure \ref{fig:square} displayed the RF metric and log-likelihood of \emph{majority-rule  consensus tree} provided by PGs with the RDouP proposal. With a higher value of $K$ (and lower number of MCMC iterations), the log-likelihood of \emph{majority-rule  consensus tree} increases and the RF metric decreases. This indicates that with a fixed computational cost, high values of $K$ are more important than the number of PGs iterations.

\begin{figure}[htp]
\centering
\includegraphics[width=0.8\textwidth]{PGS}
\includegraphics[width=0.8\textwidth]{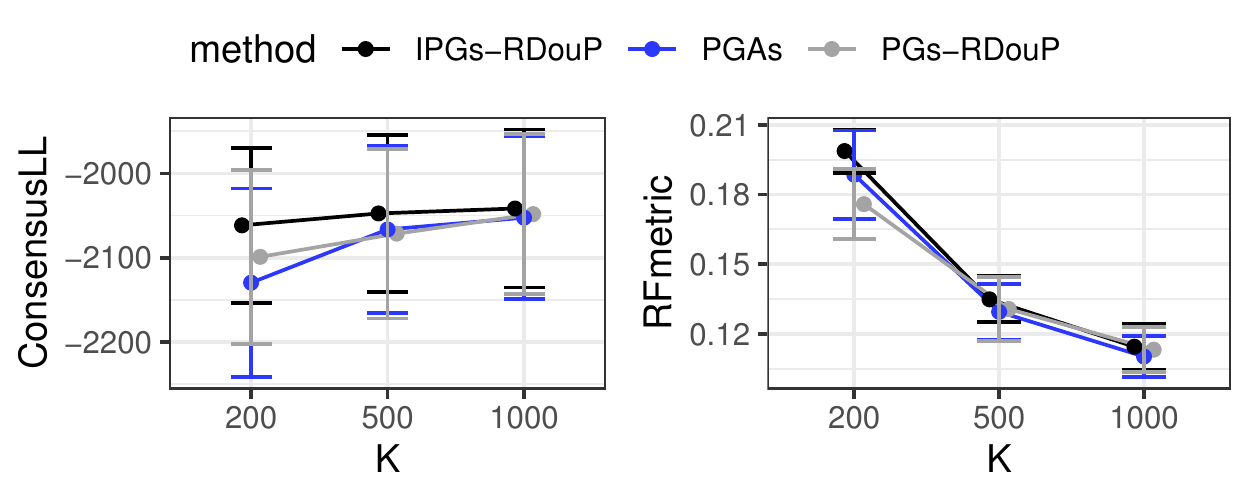}
\caption{Comparison of PGs and IPMCMC with the vanilla version of  CSMC (IPGS, PGS) and CSMC-RDouP  (IPGS-RDouP, PGS-RDouP), PGAs with CSMC-RDouP  (PGAs) as a function of number of particles. The x-axis represents number of particles. The three levels from left to right are $200$, $500$, $1000$,  respectively. The y-axis of represents the log likelihood of the consensus tree  and Robinson Foulds metric. The number of MCMC iterations are fixed at $4000$. The lower panel is a zoomed-in of the upper panel to better show the results with the RDouP proposal.} 
\label{fig:PGS-SM}
\end{figure}

We conducted another experiment to investigate the performance of  the vanilla CSMC and CSMC-RDouP  in PGs and IPMCMC,  as a function of the number of particles. We also investigated the performance of PGAs with CSMC-RDouP. The initial reference trajectory $\aus_{1:R}^*$ was simulated through the proposal distribution of the CSMC algorithm. The initial value of $\theta$ was randomly sampled from $Unif(1, 3)$.  We simulated $5$ clock trees with $15$ leaves and one dataset for each tree. The number of nucleotides on each leaf is $500$. 
Let $M$ denote the total number of nodes for running both conditional CSMC and CSMC algorithms and $P$ for the number of nodes running conditional CSMC. We set $P=2$ and $M = 2P$. The computational cost of IPMCMC is a linear function of $M$. In order to fix the computational budget for IPMCMC and PGs, we set $K$ for PGs and PGAs to be $4$ times as large as that of IPMCMC. We also set $K$ for the vanilla version of  CSMC to be three times as large as CSMC-RDouP. The budget we allocated is in favour for the vanilla CSMC. The other setups are the same as for the previous experiment. We chose three levels of $K$ for PGs with the  CSMC-RDouP: $K = 200, 500, 1000$. The number of MCMC iterations was set to $4000$ iterations, which is large enough to get convergence. 
Figure \ref{fig:PGS-SM} displays the comparison of PGs and IPMCMC with vanilla version of  CSMC (IPGS, PGS) and CSMC-RDouP  (IPGS-RDouP, PGS-RDouP) as a function of number of particles. 
We generated another two panels in the second row to provide a better comparison between PGs, IPMCMC, and PGAs with RDouP proposals as the results displayed in the upper panels cannot clearly show the difference between these methods. 
For both PGs and IPMCMC with the vanilla CSMC, the log-likelihood of majority-rule  consensus tree and RF metric  do not improve if we increase $K$. PGs, IPMCMC, and PGAs with CSMC-RDouP  perform better in terms of the log-likelihood and RF metric. If we increased $K$, the log-likelihoods increase and RF metrics decrease. The log-likelihood and RF metric provided by PGs, IPMCMC, and PGAs with CSMC-RDouP  are close.

\section{Real Data Analysis}
\subsection{Consensus tree of the primate dataset}
\begin{figure}[hbt!]
\centering
\includegraphics[width=0.6\textwidth]{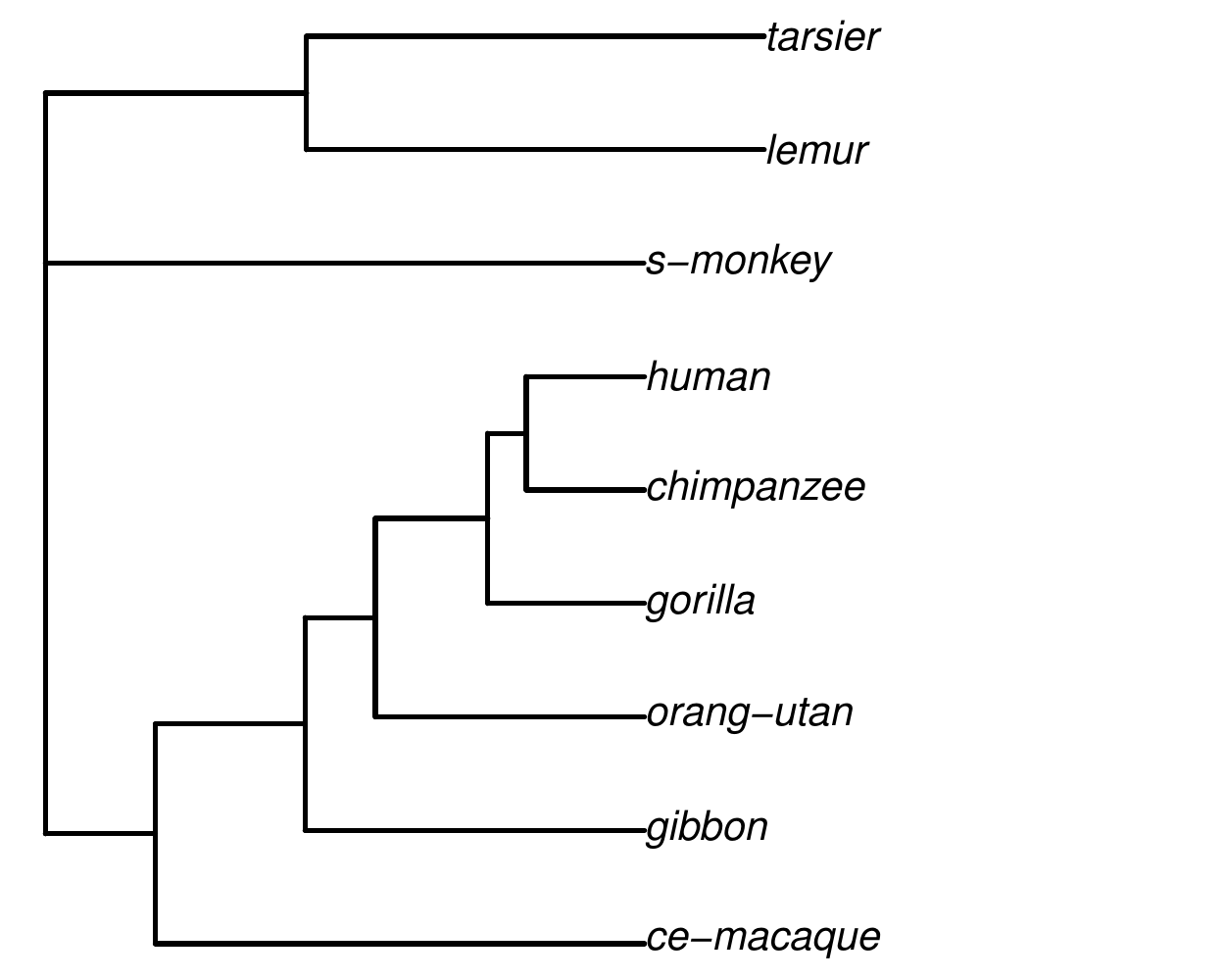}
\caption{Consensus tree of the primate dataset provided by PGs-RDouP, IPMCMC-RDouP, PGAs and MrBayes.}
\label{fig:primate}
\end{figure}

Figure \ref{fig:primate} shows the consensus tree of the primate dataset provided by PGs-RDouP, IPMCMC-RDouP, PGAs and MrBayes.

\subsection{Cichlid Fishes dataset}
The second dataset we analyze is the African cichlid fish data \citep{kocher1995, Cheon200894}, which consists of $12$ species from two tribes (\emph{Ectodini} and \emph{Lamprologini}). We compare the performance of PGs, PGs-RDouP, IPMCMC-RDouP, and PGAs with a fixed computational budget. {\color{black} The number of MCMC iterations is $N = 5000$, and the number of particles is $K=10000$. The setup for IPMCMC-RDouP is the same as \emph{Primate dataset} ($K = 2500$ in each worker of IPMCMC-RDouP). Each algorithm is repeated $12$ times. 
{\color{black} Figure 7 of \emph{Supplementary Section 5}. displays the majority-rule consensus trees provided by PGs and PGAs. The mean (standard deviation) log-likelihood for the consensus trees are showed in Table  \ref{tab:fish}.} 
The consensus tree provided by PGs-RDouP and IPMCMC-RDouP is consistent with PGAs. 
The mean and standard deviation for the posterior mean of $\theta$ for $12$ replications provided by PGAs are $8.52$ and $0.058$, respectively.


\begin{table}[h!]
  \caption{Log-likelihood of the consensus trees provided by PGs, PGs-RDou, PGAs and IPMCMC-RDouP, with mean and standard deviation.}
    \label{tab:fish}
\begin{center}
{\small
 \begin{tabular}{c |c |c |c |c} 
 &PGs &PGs-RDouP  & PGAs &IPMCMC-RDouP\\ 
 \hline
 & -6420.11  (342.8)  &  -4719.6 (12.4)  &  -4727.0 (29.5)  & -4733.8 (0.6) \\ 
\end{tabular}\vspace{0.2em}
}
\end{center}
\end{table}
}

\begin{figure}
\centering
\includegraphics[width=0.7\textwidth]{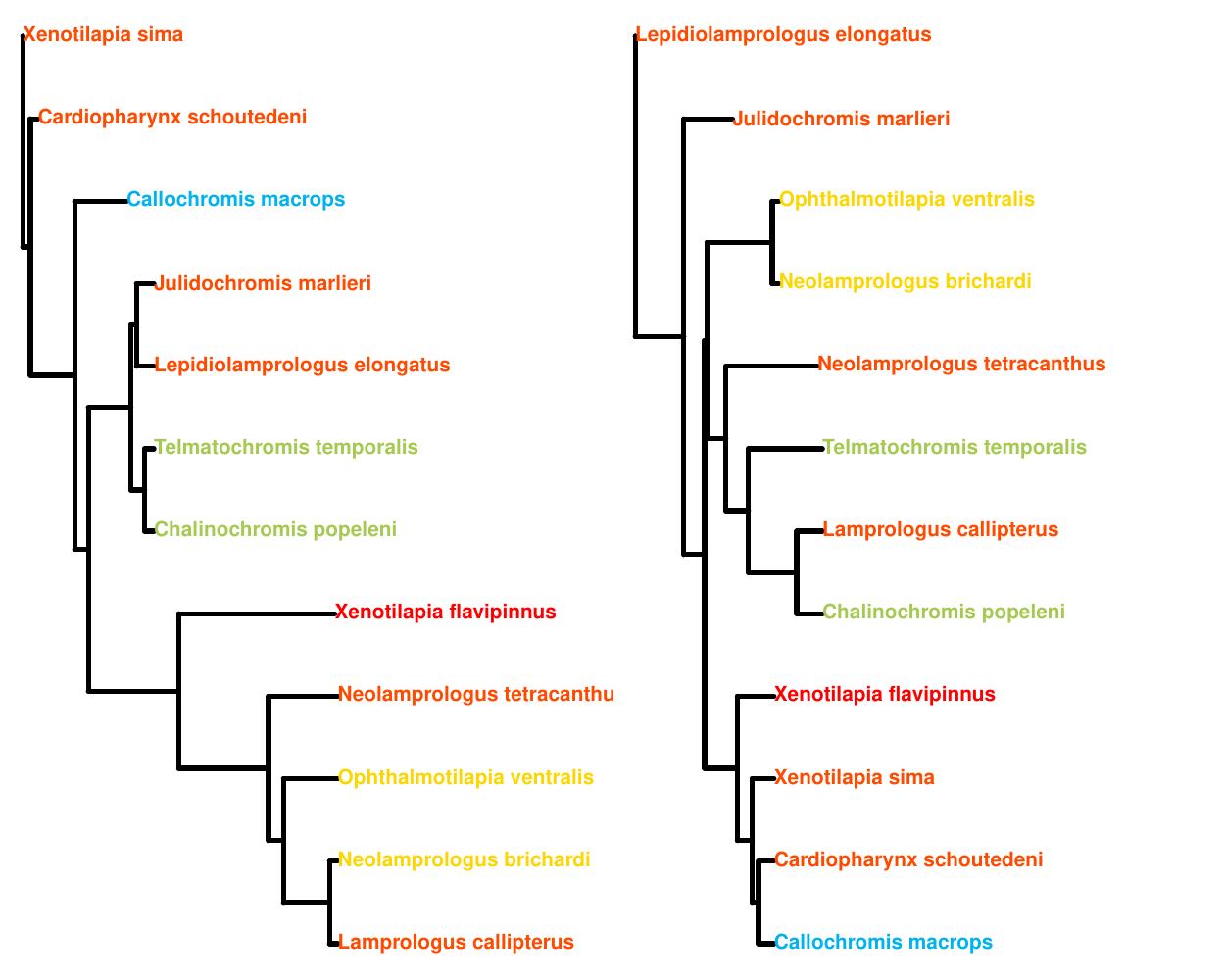}
\caption{Consensus tree of the Cichlid Fishes dataset provided by PGs (left) and PGAs (right).}
\label{fig:cfish}
\end{figure}

\begin{table*}[htb]
\caption{List of Symbols}
 \begin{tabular}{cl}\\\hline
   $y$ & observed biological sequence  \\
    $X$ &  a set of observed taxa  \\
    $t$ & a phylogenetic X-tree\\ 
    $\theta$ & parameter in the nucleotide substitution model\\
    $p(\theta)$ & the prior distribution of $\theta$\\
    $p(t)$ & the prior distribution of $t$\\
    $p(y|\theta)$ &  the likelihood of data \\
    $\pi(\cdot)$ & normalized posterior distribution of $\cdot$\\
     $\gamma(\cdot)$ & unnormalized posterior distribution of $\cdot$\\
     $q(\cdot)$ & proposal distribution for $\cdot$ \\
     $\alpha$ & MH acceptance ratio\\
     $\kappa$ & ratio of transition to transversion in K2P model \\
     $k$ & index of particle  \\
     $K$ & total number of particles \\
     $r$ & index of rank  \\
     $R$ & largest rank value \\
     $i$ & index of MCMC iteration  \\
     $N$ & total number of MCMC iterations \\
     $\bs_{r}$ & partial state with rank $r$\\
     {\color{black} $\beta(\aus)$} & {\color{black} base partial state reduced from an augmented partial state}\\
     $s_{r}$ & {\color{black} augmented} partial state with rank $r$\\
     $s_{r, k}$ & $k$-th particle for {\color{black} augmented} partial state with rank $r$\\
     $\tilde{s}_{r, k}$ & $k$-th particle for {\color{black} augmented} partial state of rank $r$ after resampling\\
     $|s|$& number of trees in forest \\
     $\varrho(\cdot)$ & the reverted {\color{black} base} state of $\cdot$\\
     $\sls(\cdot)$ & the merge state of $\cdot$\\
     $w_{r, k}$ & unnormalized weights for $k$-th particle of partial state rank $r$\\
      $W_{r, k}$ & normalized version of $w_{r, k}$\\
     $\nu^-$& a backward kernel \\
      $\nu^+$ & a forward kernel \\
      $m^-$& the backward kernel for a merge proposal\\
      $m^+$& the forward kernel for a merge proposal\\
      {\color{black} $\check{m}$} &  {\color{black} the forward kernel for a merge proposal of augmented state}\\
            $r^-$& the backward kernel for a revert step\\
      $r^+$& the forward kernel for a revert step\\
      rESS & relative effective sample size\\
      $a_{r}$ & index of ancestor sampling trajectory for partial state of rank $r$ \\
            $m$ & index of node in IPMCMC\\
      $M$ &total number of nodes in IPMCMC\\\hline
  \end{tabular}
\end{table*}

\newpage

\bibliographystyle{plain}
%


%% file: macros-topic.tex
\usepackage{amsmath, amsfonts, latexsym, amsthm, amssymb,mathrsfs,txfonts}
\usepackage{epsfig}

\newcounter{xxx}
\newtheorem{theorem}{Theorem}

\newtheorem{proposition}[theorem]{Proposition}
\newtheorem{assumption}[theorem]{Assumption}

\newcommand{\ud}{\,\mathrm{d}}

\newcommand\poset{\mathcal S}